\documentclass[twocolumn,nofootinbib,showpacs,superscriptaddress,amsmath,amssymb]{revtex4-1}

\usepackage{subfigure}
\usepackage{enumitem}
\usepackage{color}
\usepackage{graphicx}
\usepackage{dcolumn}
\usepackage{bm}
\usepackage{multirow}
\usepackage[breaklinks=true]{hyperref}
\usepackage{natbib}
\usepackage{bbm}
\usepackage{soul}
\usepackage{breakcites}

\definecolor{orange}{rgb}{1,0.5,0}
\definecolor{purple}{rgb}{0.50196078431,0,0.50196078431}
\definecolor{pink}{rgb}{1,0.07843137254,0.57647058823}
\definecolor{gray}{rgb}{0.74509803921,0.74509803921,0.74509803921}

\newcommand{\be}{\begin{equation}}
\newcommand{\ee}{\end{equation}}
\newcommand{\bea}{\begin{eqnarray}}
\newcommand{\eea}{\end{eqnarray}}

\DeclareMathOperator{\Tr}{Tr}

\begin{document}

\title{Stability of fixed points and generalized critical behavior in multifield models}

\author{A.~Eichhorn}
\email{aeichhorn@perimeterinstitute.ca}
\affiliation{
 Perimeter Institute for Theoretical Physics, 31 Caroline Street N, Waterloo, N2L 2Y5, Ontario, Canada
}
\author{D.~Mesterh\'azy}
\email{mesterh@uic.edu}
\affiliation{
 Physics Department, University of Illinois at Chicago, 845 W Taylor Street, Chicago IL 60607, USA
}
\author{M.\,M.~Scherer}
\email{scherer@thphys.uni-heidelberg.de}
\affiliation{
 Institut f\"ur Theoretische Physik, Universit\"at Heidelberg, Philosophenweg 16, 69120 Heidelberg, Germany
}

\date{\today}

\begin{abstract}
We study models with three coupled vector fields characterized by $O(N_1)\oplus O(N_2) \oplus O(N_3)$ symmetry. Using the nonperturbative functional renormalization group, we derive $\beta$ functions for the couplings and anomalous dimensions in $d$ dimensions. Specializing to the case of three dimensions, we explore interacting fixed points that generalize the $O(N)$ Wilson-Fisher fixed point. We find a symmetry-enhanced isotropic fixed point, a large class of fixed points with partial symmetry enhancement, as well as partially and fully decoupled fixed point solutions. We discuss their stability properties for all values of $N_1, N_2$, and $N_3$, emphasizing important differences to the related two-field models. For small numbers of field components we find no stable fixed point solutions, and we argue that this can be attributed to the presence of a large class of possible (mixed) couplings in the three-field and multifield models. Furthermore, we contrast different mechanisms for stability interchange between fixed points in the case of the two- and three-field models, which generically proceed through fixed-point collisions.
\end{abstract}
 

\maketitle



\section{Introduction}
\label{Sec:Introduction}

The $N$-vector model with $O(N)$ group symmetry plays an important role in the understanding of crucial aspects of renormalization group (RG) flows: In four dimensions, it exhibits a Landau pole and corresponds to a trivial theory \cite{Aizenman:1981du,Frohlich:1982tw,Luscher:1987ay,*Luscher:1987ek,*Luscher:1988uq,Rosten:2008ts}. In other words, as an interacting model it is only valid over a finite range of scales, thus constituting an effective low-energy theory. This could affect the possible range of validity of the standard model of particle physics \cite{Dashen:1983ts,Callaway:1983zd,Kuti:1987nr,Luscher:1988gc,Kogut:1988sf,Gockeler:1997dn} and could also play a role in cosmology as, e.g., many inflationary models probably share this feature. On the other hand, in three dimensions the theory exhibits an important example of an interacting RG fixed point \cite{LeGuillou:1977ju,Guida:1998bx}. Such fixed points are crucial in the understanding of scaling and universality in critical phenomena \cite{Wilson:1973jj,Wegner:1976bk,Pelissetto:2000ek} and, more recently, they have been of considerable interest, e.g., in the problem of the ultraviolet (UV) completion of gravity \cite{Weinberg:1980gg, Reuter:2012id}. On a more technical level, well-known examples such as, e.g., the infrared (IR) attractive Wilson-Fisher fixed point (FP) in the $O(N)$ model may provide an important benchmark test for nonperturbative methods, which one may then apply to other problems of interest (see, e.g., Ref.\ \cite{Delamotte:2004zg}).

\begin{figure}[!h]
\includegraphics[width=0.7\linewidth]{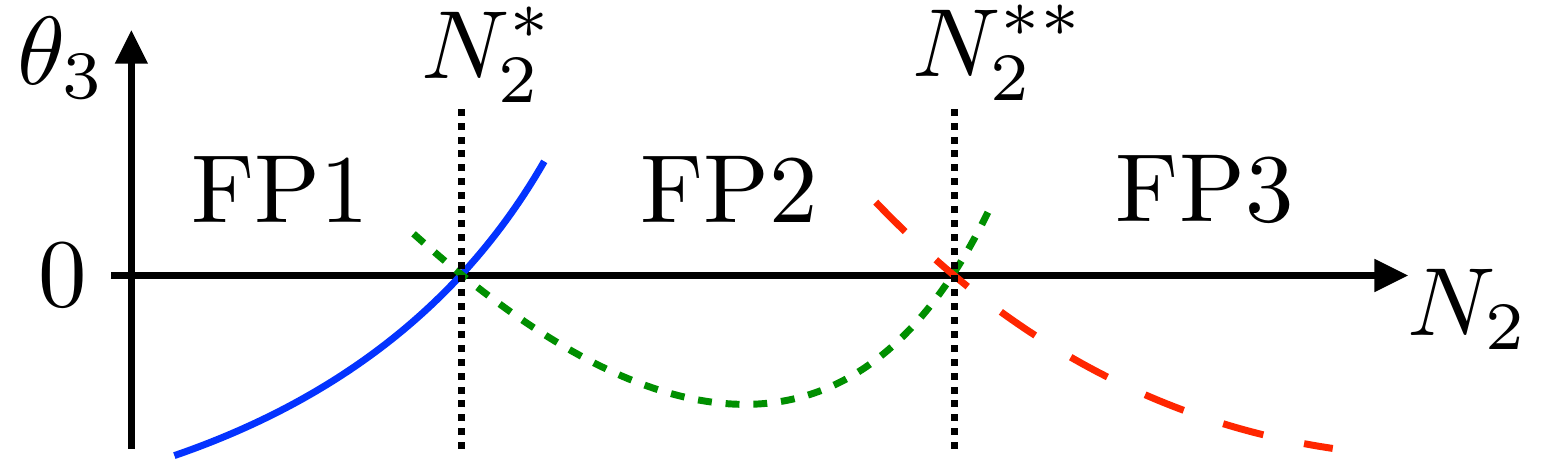}
\caption{\label{Fig:collision}We show a sketch of the value of the third largest critical exponent, $\theta_3$, of three different FPs (solid, dotted, and dashed lines) in the $O(N_1)\oplus O(N_2)$ model at fixed $N_1$ as a function of $N_2$. The regime where a FP is stable is indicated by the labels FP1, FP2 and FP3. As the coordinates of two different FPs coincide at $N_2^{\ast}$ and at $N_2^{\ast\ast}$, these FPs exchange their stability properties, and $\theta_{3}$ changes its sign if evaluated beyond the stable regime. An explicit calculation showing this situation can be found in Ref.\ \cite{Eichhorn:2013zza}.}
\end{figure}

Extending the $O(N_1)$ vector model by a coupling to another $O(N_2)$ symmetric vector field leads to complex dynamics that has been discussed extensively in the context of multicritical phenomena and systems with competing order parameters \cite{Fisher:1974zz,Kosterlitz:1976zza,Aharony:2002,Aharony:2002zz,Calabrese:2002bm,Folk:2008mi}. Such a theory is characterized by an $O(N_{1})\oplus O(N_{2})$ symmetry which admits a number of interacting (IR attractive) FPs. These travel through the coupling space of the model as the numbers of field components $N_{1}$ and $N_{2}$ are varied. At particular values of $N_1$ and $N_2$, two of these FPs can collide and exchange their stability properties, cf.\ Fig.\ \ref{Fig:collision}. In this context, an IR stable FP is defined as featuring only two positive critical exponents, as this corresponds to the number of relevant couplings that need to be tuned in order to approach the FP. When two FPs collide, a FP with three positive critical exponents trades one of them for a negative exponent, while the second FP picks up the additional relevant direction and becomes unstable. As a consequence, it turns out that for every combination of $N_1$ and $N_2$ there is exactly one stable FP. Of course, this statement assumes that one considers renormalization group trajectories within a single domain of attraction. In general, the parameter/coupling space of the model will allow for separate domains, where different FPs might exist, and may or may not be stable. 

In this work, we will for the first time provide a comprehensive analysis of a model with a coupling to an additional field, with a resulting $O(N_1) \oplus O(N_2) \oplus O(N_3)$ symmetry. Note that the $O(N_1) \oplus O(N_2)$ symmetry acts trivially on the third vector, and similarly the first two fields transform as singlets under the $O(N_3)$ symmetry. At first, one might expect that this model will exhibit very similar behavior to that already encountered in the case of the two-field models, and will feature a single stable FP with three relevant directions at every value of $(N_1, N_2, N_3)$. In this study, we present evidence for a rather different behavior, where FPs exhibit a large number of relevant eigendirections in a given range of values for the number of field components. This leads to the absence of stable FP solutions in a part of the parameter space. In the following, we will argue that this is a generic feature of multifield models and is due to a significantly increased number of possible mixed interactions compared to the single-field or two-field models. This behavior is akin to the absence of FPs in the low-energy effective models for phases of strongly-interacting matter \cite{Basile:2005hw,Vicari:2007ma}, or frustrated spin systems \cite{Pelissetto:2000ne,Tissier:2002zz}. In both cases, one observes the absence of stable FP solutions beyond some critical number of field components, indicative of a first order phase transition (see, e.g., Ref.\ \cite{Bak:1976}).

The presence of competing orders can change the nature of a continuous transition or even drive it to be first order. Although there are numerous examples in the literature, this is probably best illustrated using the example of the two-field model \cite{Fisher:1974zz,Kosterlitz:1976zza,Aharony:2002,Aharony:2002zz,Calabrese:2002bm,Folk:2008mi}. Already at the mean-field level one observes a suppression of the coexistence region in the presence of a strong repulsive interaction between the two competing fields \cite{She:2010}. If the coupling assumes a critical value, the coexistence region vanishes and the second-order lines merge to a first order transition. This dramatic change in the phase diagram marks a change in the universality class of the adjacent multicritical point. Fluctuations will favor either one of these scenarios (corresponding to a tetracritical or bicritical point) as long as the associated fixed point is within its domain of attraction.\footnote{This is not the case if the interactions are sufficiently strong and the total number of field components $N = N_{1} + N_{2}$ is larger than some critical value. Then, instead of a multicritical point one observes a genuine first order transition in the phase diagram of the model.} Multifield models feature a different possibility: fluctuation-mediated interactions might not only affect the universality class of the multicritical point, but they might even render it unstable, thus allowing no IR convergent RG trajectory. The difference between two- and multifield models lies in the distinct RG flow topologies and mechanisms of stability trading between different fixed points, which leads to the absence of a stable fixed point. We will investigate these properties in detail in the following sections, as well as the general behavior of systems with a large number of interacting sectors.

The main motivation for this study is to gain an improved understanding of generalized Wilson-Fisher universality classes (multicritical points) and to understand how these might show up in the phase diagram of systems with multiple order parameters. Previous work, in the context of systems with two competing order parameters, has led to the understanding that the effect of fluctuations plays an important role in addressing the stability of a particular continuous transition \cite{Fisher:1974zz,Kosterlitz:1976zza,Aharony:2002,Aharony:2002zz,Calabrese:2002bm,Folk:2008mi}. The $SO(5)$ theory of high-$T_c$ superconductivity \cite{Calabrese:2002bm,Demler:2004} provides one example, where fluctuations of the order parameters are seen to alter the stability properties of the associated fixed point and rule out such a theory as an effective IR description in the region where both order parameters become critical. On the other hand, it has been pointed out that the interplay of two competing order parameters might explain the presence of first order transitions or spatially inhomogeneous phases that exhibit finite wavevector ordering near quantum criticality \cite{She:2010}. Here, we argue that first order transition might be a generic scenario for systems that feature a large number of competing phases.

The outline of this paper is as follows: In Sec.~\ref{Sec:Model} we present the model under consideration in detail. In Sec.~\ref{Sec:Fixed-point analysis} we then explain the results of our study, discussing numerical results and scaling relations for several different FPs. In the Appendix we present the renormalization group flow equations for these models in $d$ dimensions, both in a local potential approximation (LPA) and including anomalous dimensions. Sections \ref{Sec:Model} and \ref{Sec:Fixed-point analysis} are self-contained, and can be read without referring to the technical details of our study.


\section{Model}
\label{Sec:Model}

We consider a model with three different bosonic fields, $\phi_1,\phi_2$, and $\phi_3$, with $N_1,N_2$, and $N_3$ field components, respectively. We derive the $\beta$ functions from the nonperturbative functional flow equation for the (Euclidean) scale dependent effective action $\Gamma_{k}$ \cite{Wetterich:1992yh}, (see Appendix for details, and reviews, e.g., Ref.\ \cite{Berges:2000ew,Polonyi:2001se,Pawlowski:2005xe,Gies:2006wv,Delamotte:2007pf,Rosten:2010vm,Metzner:2011cw}). This method has been shown to yield results in very good agreement with those obtained from the $\epsilon$-expansion and lattice simulations in the case of the $O(N)$ Wilson-Fisher FP, see, e.g., Ref.\ \cite{Canet:2003qd, Bervillier:2007rc,Litim:2010tt,Codello:2012ec} and the $O(N_1)\oplus O(N_2)$ FPs \cite{Eichhorn:2013zza}. To leading order in the derivative expansion \cite{Morris:1994au, *Morris:1994ie, *Morris:1996kn, *Morris:1996xq} our \textit{ansatz} for $\Gamma_{k}$ reads
\bea
\Gamma_{k} &=& \int\! d^dx \left( \sum_{I=1}^3 Z_{I} \left( \partial_{\mu} \phi_{I} \right)^{2} + U_k (\phi_{1}, \phi_{2}, \phi_{3})\right) ~, 
\label{Eq:EffectiveAction}
\eea
where $\phi_{I}^{a}$, $a = 1, \ldots, N_{I}$, and $\phi_{I}^{2} \equiv \phi_{I}^{a} \phi_{I}^{a}$. Here, we have introduced the scale dependent effective potential
\be
\hspace{-5pt} U_k = \sum_{l,m,n} \frac{\bar{\lambda}_{l,m,n}}{l!\, m!\, n!} \left( \bar{\rho}_{I} - \bar{\kappa}_{1} \right)^{l} \left( \bar{\rho}_{2} - \bar{\kappa}_{2} \right)^{m} \left( \bar{\rho}_{3} - \bar{\kappa}_{3} \right)^n , 
\label{Eq:EffectivePotential}
\ee
which we have written in terms of the invariants $\bar{\rho}_{I} = \frac{1}{2} \phi_{I}^{2}$, thereby making the $O(N_1) \oplus O(N_2) \oplus O(N_3)$ symmetry manifest. The parameter $k$ defines an infrared momentum cutoff scale, on which the parameters and couplings depend. For brevity we do not indicate the scale dependence explicitly, i.e., $\bar{\lambda}_{l,m,n} = \bar{\lambda}_{l,m,n} (k)$. Similarly, scale dependent wavefunction renormalization factors are simply denoted by $Z_{I}$. We expand the scale dependent effective potential $U_{k}$ around (possibly) nonvanishing scale dependent minima for the fields, $\bar\kappa_{I}$.

For the identification of scaling solutions, we introduce dimensionless renormalized couplings, given by
\bea
u_k &=& U_k k^{-d} ~, \nonumber\\ 
\kappa_{I} &=& Z_I k^{2-d}\bar{\kappa}_{I} ~, \nonumber\\ 
\rho_{I} &=& {Z_I k^{2-d}} \bar{\rho}_{I} ~,\nonumber\\
\lambda_{l,m,n} &=& \bar{\lambda}_{l,m,n} Z_{1}^{-l}\, Z_{2}^{-m}\, Z_{3}^{-n} k^{-d + (l+m+n) (d-2)} ~.
\eea
We truncate the coupling space to a finite-dimensional subspace of the form Eqs.\ \eqref{Eq:EffectiveAction} and \eqref{Eq:EffectivePotential}, which includes all relevant operators, i.e., those with a positive critical exponent at the FP of interest. Including field monomials up to order $4$, $6$, and $8$, defines the local potential approximation, LPA 4/$4+\eta$, LPA 6/$6+\eta$, and the LPA 8/$8+\eta$, respectively (depending on the inclusion of a scale dependent wavefunction renormalization, $\partial_{t} Z_{I} \neq 0$).

In order to distinguish physically meaningful from spurious FPs arising within a given truncation, we demand that a FP can be continued to higher orders in the truncation, and universal quantities, e.g., critical exponents, show signs of convergence. Further, corrections to canonical scaling should not be too large, as otherwise we would not expect our truncation to be reliable. Moreover, we demand that all eigenvalues of $\left. \left(\frac{\partial^2 u_k}{\partial \rho_I \partial \rho_J} \right)\right|_{\rho_I = \kappa_I}$ are non-negative. If this condition is violated, the expansion point for the effective potential does not correspond to its true minimum, and critical exponents evaluated around this point will show poor convergence properties. The parameter $\Delta \equiv \left. \det \left(\frac{\partial^2 u_k}{\partial \rho_I \partial \rho_J} \right) \right|_{\rho_I = \kappa_I}$ serves to separate the space of couplings into different (not necessarily bounded) domains of attraction. Within such a domain there exists at most \textit{one} IR stable FP, characterized by the strength of correlations \cite{Vicari:2006xr}. In the following, we will be interested specifically in IR scaling solutions in the $\Delta \geq 0$ domain, corresponding to a minimum of the effective potential.

With these preliminaries and definitions we now turn to analyze the fixed-point structure of this model.


\section{Fixed-point analysis}
\label{Sec:Fixed-point analysis}

For generic multifield models with $\bigoplus_{I} O(N_{I})$ symmetry, a number of FPs and their stability properties can be deduced from the existence of the $O(N)$ Wilson-Fisher FP. These FPs are typically characterized by an enhancement of symmetry.

\begin{itemize}[leftmargin=*]

\item[] The \textit{isotropic fixed point (IFP)} shows maximal symmetry enhancement: All couplings at a given order in the fields take the same value, i.e., in the three-field model we have $\left. \lambda_{l,m,n} \right|_{l+m+n = 2} \equiv \lambda_{2}$, and similarly for higher order couplings. It is characterized by $O(N)$ symmetry, where $N \equiv \sum N_{I}$. Accordingly, it features additional massless Goldstone modes, even in the case of an underlying discrete symmetry, e.g., with $Z_{2} \oplus Z_{2} \oplus Z_{2}$ symmetry.

\item[] The \textit{decoupled fixed point (DFP)} is characterized by vanishing couplings between different sectors of the theory. In a model with three fields this implies $\lambda_{l,m,n} = 0$ if $l,m \neq 0$, $l,n \neq 0$, or $m,n \neq 0$. The values of the couplings in each sector approach those of the corresponding $O(N_{I})$ Wilson-Fisher FP. However, while the action at that FP is fully decoupled, critical exponents that relate to mixed couplings are nontrivial.

\item[] The \textit{decoupled isotropic fixed point (DIFP)} occurs for the first time in a model with three fields: It is characterized by a partial enhancement of symmetry, as two fields remain fully coupled and the couplings in those sectors become degenerate. Simultaneously, the third field decouples completely and its couplings approach the corresponding values of the Wilson-Fisher FP. There exist three realizations of this FP, as any of the three sectors can be the one to decouple. For generic multifield models, a set of different DIFPs exists, where any number of the fields decouple, and the couplings in the remaining sectors show a symmetry enhancement.

\item[] We may additionally infer the existence of another class of FPs from the knowledge of the anisotropic scaling solution in the two-field model with $O(N_{1})\oplus O(N_{2})$ symmetry. In general, any FP of the two-field model can be extended to the three-field model as a partially decoupled FP, where the third field decouples from the other two and the fixed-point values of its couplings are given by those of the Wilson-Fisher fixed point. In particular, this applies to the \textit{biconical} fixed-point solution in the two-field model, which we identify as the \textit{decoupled biconical fixed point (DBFP)}.

\end{itemize}

In analogy to the two-field case, we will refer to a FP in the three-field model as stable, if it features three relevant directions. This terminology relates to the requirement that these three parameters need to be tuned to reach the multicritical point.

Our model, cf.\ Eqs.\ \eqref{Eq:EffectiveAction} and \eqref{Eq:EffectivePotential}, contains nine running couplings in the potential when we take into account all operators up to fourth order in the fields, that is, the associated parameters $\kappa_I$ and couplings $\lambda_{l,m,n}$ with $l + m + n = 2$. They give rise to the nine largest critical exponents of the model\footnote{These can be calculated from the stability matrix
\be
\mathit{\Theta}_{i,j} = \left. \frac{\partial \beta_{g_i}}{\partial g_j} \right|_{\textrm{FP}} ~.
\ee
Here, the $g_i$ label all the (dimensionless renormalized) running couplings/parameters and $\beta_{g_i}$ define the corresponding beta functions. The critical exponents are then given by:
\be
\theta_{i} ~ \in - {\rm spec} (\mathit{\Theta}) ~.
\ee
}.
Going to higher orders in the expansion of the effective potential, the number of running couplings $\lambda_{l,m,n}$ increases. Accordingly, the number of critical exponents will increase, but those subleading critical exponents will be irrelevant. The leading order critical exponents will typically receive corrections from the additional higher-order couplings and will therefore vary with the order of the truncation. 


\subsection{Isotropic fixed point}
\label{SubSec:Isotropic fixed point}

\begin{figure}[!t]
\includegraphics[width=0.7\linewidth, height=120pt, clip=true, trim= 10cm 8cm 0cm 0cm]{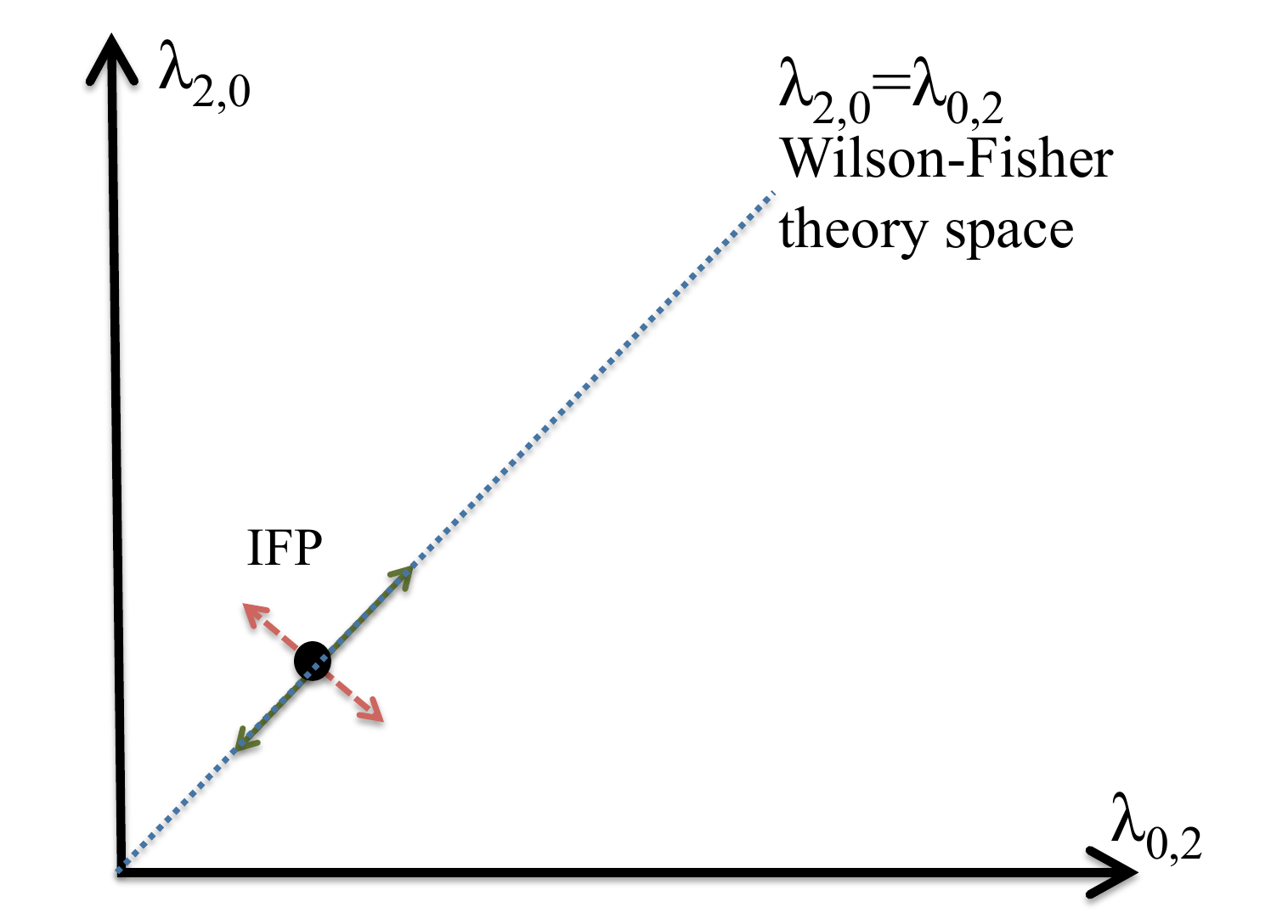}
\caption{\label{IFP_illustration}This illustration of the two-dimensional subspace of the two-field coupling space shows that the Wilson-Fisher theory space is a one-dimensional subspace, corresponding to the $\lambda_{2,0} = \lambda_{0,2}$ line. The critical exponent along this direction corresponds to the largest Wilson-Fisher critical exponent. Another eigendirection of the stability matrix does not respect the enhanced symmetry of the FP.}
\end{figure}

To determine the critical exponents at the IFP in multifield models, it is crucial to realize that a subset of those is determined by the $O(N)$ Wilson-Fisher exponents. These correspond to the directions in theory space that respect that full symmetry, i.e., those directions that span the Wilson-Fisher theory space. Additional directions in the full theory space break (a subgroup) of the enhanced $O(N)$ symmetry, and their associated critical exponents are therefore not associated with the Wilson-Fisher critical exponents, see Fig.~\ref{IFP_illustration}. 

It accordingly follows, that of the nine largest critical exponents at the IFP, two are determined by the scaling exponents of the $O(N)$ Wilson-Fisher FP. Among the additional critical exponents, one can observe a degeneracy, which can be understood from the following considerations:

\begin{table*}[!t]
\renewcommand{\arraystretch}{1.2}
\renewcommand{\tabcolsep}{6pt}
\begin{tabular}{c | c c c c c}
$N = \sum N_{I}$ & $ \theta_{1} = \theta_{2} = y_{2,2}$ & $\theta_{3} = \frac{1}{\nu}$ & $\theta_{4} = \theta_{5} = \theta_{6} = y_{4,4}$ & $\theta_{7} = \theta_{8} = y_{4,2}$ & $\theta_{9} = -\omega$ \\
\hline\hline
3 & 1.790 & 1.362 & 0.086 & -0.380 & -0.756 \\
4 & 1.818 & 1.292 & 0.196 & -0.324 & -0.775 \\
5 & 1.842 & 1.240 & 0.289 & -0.283 &-0.797 \\ 
\hline\hline
\end{tabular}
\caption{\label{Tab:IFP}Critical indices for the IFP in LPA 12 including anomalous dimensions. Our notation corresponds to the one introduced in Ref.\ \cite{Eichhorn:2013zza}.}
\end{table*}

\begin{figure}[!t]
\includegraphics[width=0.8\linewidth]{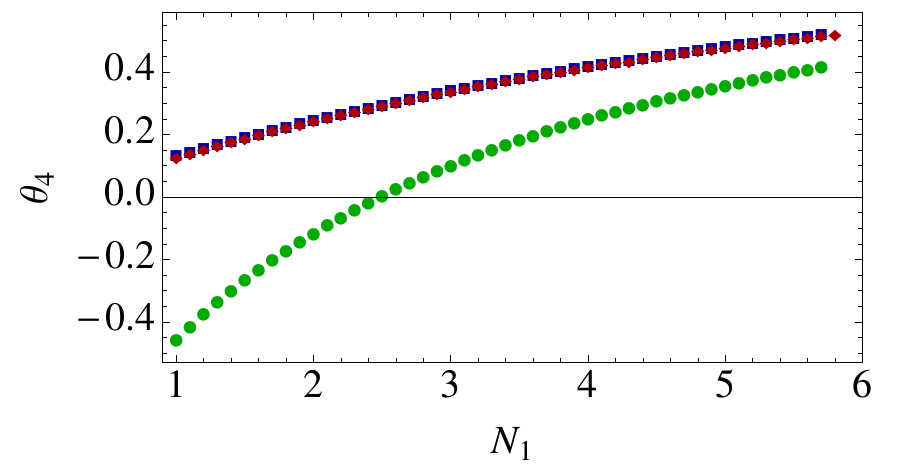}\\
\includegraphics[width=0.8\linewidth]{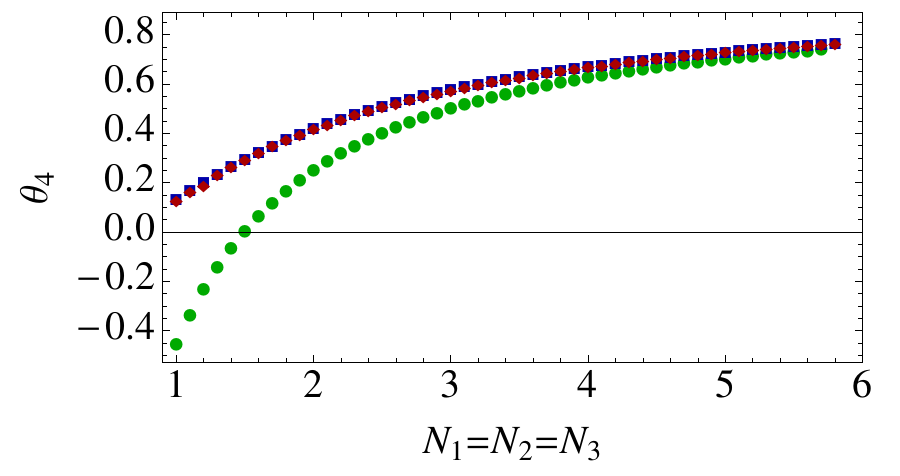}
\caption{\label{Fig:IFP}(Color online) We show the fourth-largest critical exponent at the IFP as a function of $N_1$, with $N_2=N_3=1$ (upper panel) and as a function of $N_{1} = N_{2} = N_{3}$ (lower panel), with the LPA 4 result (green circles), LPA 6 (blue squares) and LPA 8 (red diamonds). The LPA clearly converges rapidly. As this exponent is positive beyond the LPA 4 and shows a three-fold degeneracy, the IFP is characterized by a total of six relevant directions.}
\end{figure}

Sufficiently close to $d = 4$, i.e., in the vicinity of the noninteracting fixed point, the relevant perturbations at the Wilson-Fisher FP are determined by the spin-$l$ representations of the $O(N)$ symmetry group \cite{Wegner:1972zz} (see also Ref.\ \cite{Calabrese:2002bm}). Here, we will assume that such a classification of perturbations also holds for arbitrary dimensions, and only operators up to quartic order need to be taken into account. Defining the $N$-component field $\Phi=(\phi_1,\phi_2,\phi_3)$, we find the following eigendirections of the stability matrix at the IFP in the three-field model:

\begin{itemize}[leftmargin=*]

\item[] A scalar quadratic perturbation at the IFP $\sim m^{2} \Phi^{2}$, where $P_{2,0} = \Phi^2$, defines the critical exponent $\nu$ related to the divergence of the correlation length, i.e., $[m^{2}] = \frac{1}{\nu}$. This critical exponent is thus always positive, corresponding to one relevant direction. 

\item[] From the quadratic perturbation $P^{a b}_{2,2} = \Phi^{a} \Phi^{b} - \frac{1}{N} \delta^{a b} \Phi^{2}$ in the spin-$1$ representation of the $O(N)$ symmetry group, we can construct an $O(N)$ invariant operator by a suitable contraction of indices where, e.g., $\mathcal{P}_{2,2} = \phi_{I}^{2} - \frac{N_{I}}{N} \Phi^{2}$, $I = 1,2,3$. Then, the perturbation $\sim v \mathcal{P}_{2,2} $ defines the critical exponent $y_{2,2} = [v] = d - [\mathcal{P}_{2,2}]$. For the three-field model, two independent operators of that form can be constructed. Thus the corresponding critical exponent shows a two-fold degeneracy in the scaling spectrum. We emphasize that these critical exponents are identical to those evaluated for the $O(N)$ symmetric IFP in the two-field or anisotropic $N$-vector models, see, e.g., Ref.\ \cite{Calabrese:2002bm, Folk:2008mi, 2011PhRvB..84l5136H, Eichhorn:2013zza}. These exponents are always positive, adding two further relevant directions at the FP.

\item[] A scalar quartic perturbation $\sim u \Phi^{4}$, where $P_{4,0} = \Phi^{4}$, which is irrelevant at the IFP and defines the Wegner critical exponent $\omega$, yields a negative critical exponent, i.e., $[u] = y_{4,0} = - \omega$. 

\item[] A quartic operator in the spin-$1$ representation of the $O(N)$ symmetry group: $P_{4,2}^{a b} = \Phi^{2} P_{2,2}^{a b}$ can be contracted to define the exponent $y_{4,2} = d - [\mathcal{P}_{4,2}]$, which is also given by the value calculated in the two-field model and shows a two-fold degeneracy in the three-field case.

\item[] A quartic perturbation in the spin-$2$ representation of the $O(N)$ symmetry group is given by
\bea
\hspace{8pt} P_{4,4}^{a b c d} &=& \Phi^{a} \Phi^{b} \Phi^{c} \Phi^{d} \nonumber - \frac{1}{N + 4} \Phi^{2} \left( \Phi^{a} \Phi^{b} \delta^{c d} + p(a,b,c,d) \right) \nonumber \\ && +\: \frac{1}{(N + 2) (N + 4)} ( \Phi^{2} )^{2} \left( \delta^{a b} \delta^{c d} + p(a,b,c,d) \right) ~. \nonumber\\ &&
\eea
Note, that $p(a,b,c,d)$ denotes inequivalent permutation of the indices on the preceding operator, e.g., $\Phi^{a} \Phi^{b} \delta^{c d} \rightarrow \Phi^{c} \Phi^{d} \delta^{a b} + \ldots$ or $\delta^{a b} \delta^{c d} \rightarrow \delta^{a c} \delta^{b d} + \ldots$ The corresponding perturbation defines the critical exponent $y_{4,4} = d - [\mathcal{P}_{4,4}]$, which becomes negative for $N = 3$, cf.\ Fig.\ \ref{Fig:IFP}. This critical exponent with three-fold degeneracy is again determined by the two-field model, see, e.g., Ref.\ \cite{Calabrese:2002bm, Eichhorn:2013zza}.
\end{itemize}

Our explicit numerical results within the LPA up to 8th order of the three-field model confirm this picture. We may therefore directly exploit the LPA to order 12 including anomalous dimensions within the two-field model to obtain the corresponding exponents. Note that a comparable computation in the full three-field model is quite demanding due to the large number of couplings between different sectors of the model. Using our results for the two-field case, see Ref.\ \cite{Eichhorn:2013zza}, we can accordingly determine the nine largest critical exponents of the model, cf.\ Tab.\ \ref{Tab:IFP}. In general the IFP shows a large number of relevant parameters that require tuning to approach the FP. In fact, the IFP is unstable for any integer combination of field components $(N_1,N_2,N_3)$. It is clear that this pattern will persist to generic multifield models where additional fields are coupled to the system -- for each additional sector the number of relevant directions at the IFP increases (at least) by three.


\subsection{Decoupled fixed point}
\label{SubSec:Decoupled fixed point}

At the DFP, the nonvanishing couplings, i.e., the mass parameters and self-couplings lead to one relevant and one irrelevant direction in each sector. The associated critical exponents are those of the corresponding $O(N_{I})$ Wilson-Fisher FP. While the mixed couplings such as $\lambda_{1,1,0}$ vanish at the FP, the corresponding critical exponents are nontrivial. This follows, as the FP is an interacting FP, and these residual interactions affect scaling dimensions of operators even if the corresponding coupling vanishes. In other words, contributions \mbox{$\sim \lambda_{1,1,0} \lambda_{2,0,0}$} in the $\beta$ functions yield nonvanishing entries in the stability matrix even if $\lambda_{1,1,0} = \lambda_{1,0,1} = \ldots = 0$.

\begin{table}[!t]
\renewcommand{\arraystretch}{1.4}
\renewcommand{\tabcolsep}{4pt}
\begin{tabular}{c c c | c c c c c c}
$N_1$& $N_2$&$N_3$ & $\theta_{1}$ & $\theta_{2}$ & $\theta_{3}$ & $\theta_{4}$ & $\theta_{5}$ & $\theta_{6}$ \\
\hline\hline
1& 1& 1& 1.571 & 1.571 &1.571 & 0.142 & 0.142& 0.142\\
2&1&1&1.459 & 1.571 & 1.571 & 0.030 & 0.030 & 0.142\\
3&1&1&1.367 & 1.571 & 1.571 & -0.062 & -0.062 & 0.142\\
4&1&1&1.296 & 1.571 &1.571 & -0.133& -0.133& 0.142\\
\hline
2&2&1 & 1.459 & 1.459 & 1.571 & -0.082 & 0.030& 0.030\\
3&2&1& 1.367 & 1.459 & 1.571 & -0.174 & -0.062& 0.030\\
4&2&1&1.296 & 1.459&1.571& -0.245& -0.133 & 0.030\\
3&3&1&1.367 &1.367&1.571& -0.266& -0.062 & -0.062\\
\hline
2&2&2& 1.459 & 1.459 & 1.459 &-0.082 & -0.082 & -0.082\\
3&2&2& 1.367 & 1.459 & 1.459& -0.174 & -0.174 & -0.082\\ 
\hline\hline
\end{tabular}
\caption{\label{Tab:DFPa}We list the six largest critical exponents as a function of $N_1, N_2, N_3$ at the DFP, employing results from the LPA 12 including anomalous dimensions from \cite{Eichhorn:2013zza} and using the scaling relations Eqs.\ \eqref{Eq:ScalingRelationsDFPa} -- \eqref{Eq:ScalingRelationsDFPc}.}
\end{table}

At the DFP, the eigendirections corresponding to the six largest critical exponents can be determined using a scaling relation: The quartic couplings $\lambda_{1,1,0}, \lambda_{1,0,1}$, and $\lambda_{0,1,1}$ correspond to eigendirections of the FP with critical exponents
\bea
\theta_{4} &=& \frac{1}{\nu_1} + \frac{1}{\nu_2}-d ~, \label{Eq:ScalingRelationsDFPa} \\
\theta_{5} &=& \frac{1}{\nu_1} + \frac{1}{\nu_3}-d ~, \label{Eq:ScalingRelationsDFPb} \\
\theta_{6} &=& \frac{1}{\nu_2} + \frac{1}{\nu_3}-d ~, \label{Eq:ScalingRelationsDFPc}
\eea
where $\nu_{I} = \frac{1}{\theta_{I}}$, $I = 1,2,3$. These scaling relations can be  motivated as follows \cite{Aharony:1976,Aharony:2002}: At the DFP, the decoupling of the three sectors implies $[\phi_I^{2} \phi_{J}^{2}] = [\phi_{I}^{2}] + [\phi_{J}^{2}]$. Furthermore, the scaling dimensions of $\phi_{I}^{2}$ are -- due to the decoupling -- determined by the Wilson-Fisher critical exponents, such that $[\phi_{I}^{2}] = - \frac{1}{\nu_{I}} + d$. The relations Eqs.\ \eqref{Eq:ScalingRelationsDFPa} -- \eqref{Eq:ScalingRelationsDFPc} follow directly.

\begin{table}[!t]
\renewcommand{\arraystretch}{1.4}
\renewcommand{\tabcolsep}{4pt}
\textbf{LPA 8}
\vskip 2pt
\begin{tabular}{c c c | c c c c c c}
$N_1$ & $N_2$ & $N_3$ & $\theta_{1}$ & $\theta_{2}$ & $\theta_{3}$ & $\theta_{4}$ & $\theta_{5}$ & $\theta_{6}$ \\
\hline\hline
1 & 1 & 1 & 1.537 & 1.537 & 1.537 & 0.067 & 0.067 & 0.067\\
2 & 1 & 1 & 1.399 & 1.537 & 1.537 & -0.057 & -0.057 & 0.067\\
3 & 2 & 1 & 1.306 & 1.399 & 1.537 & -0.275 & -0.150& -0.057\\
\hline \hline
 & & & \multicolumn{2}{c}{$\Delta \theta_{4}$} & \multicolumn{2}{c}{$\Delta \theta_{5}$} & \multicolumn{2}{c}{$\Delta \theta_{6}$} \\ \hline \hline
1 & 1 & 1 & \multicolumn{2}{c}{0.007} & \multicolumn{2}{c}{0.007} & \multicolumn{2}{c}{0.007}\\
2 & 1 & 1 & \multicolumn{2}{c}{-0.007} & \multicolumn{2}{c}{-0.007} & \multicolumn{2}{c}{0.007}\\
3 & 2 & 1 & \multicolumn{2}{c}{-0.020} & \multicolumn{2}{c}{-0.007} & \multicolumn{2}{c}{-0.007}\\ \hline \hline
\end{tabular}
\vskip 10pt
\textbf{LPA 8}$\mathbf{+\eta}$
\vskip 2pt
\begin{tabular}{c c c | c c c c c c}
$N_1$ & $N_2$ & $N_3$ & $\theta_{1}$ & $\theta_{2}$ & $\theta_{3}$ & $\theta_{4}$ & $\theta_{5}$ & $\theta_{6}$ \\ \hline\hline
1 & 1 & 1 & 1.564 & 1.564 & 1.564 & 0.080 & 0.080 & 0.080\\
2 & 1 & 1 & 1.447 & 1.564 & 1.564 & -0.028 & -0.028 & 0.080\\
3 & 2 & 1 & 1.359 & 1.447 & 1.564 & -0.220 & -0.112 & -0.028\\
\hline \hline
 & & & \multicolumn{2}{c}{$\Delta \theta_{4}$} & \multicolumn{2}{c}{$\Delta \theta_{5}$} & \multicolumn{2}{c}{$\Delta \theta_{6}$} \\ \hline \hline
1 & 1 & 1 & \multicolumn{2}{c}{0.047} & \multicolumn{2}{c}{0.047} & \multicolumn{2}{c}{0.047}\\
2 & 1 & 1 & \multicolumn{2}{c}{0.038} & \multicolumn{2}{c}{0.038} & \multicolumn{2}{c}{0.047}\\
3 & 2 & 1 & \multicolumn{2}{c}{0.027} & \multicolumn{2}{c}{0.035} & \multicolumn{2}{c}{0.038}\\
\hline\hline
\end{tabular}
\caption{\label{Tab:DFPb}We list the six largest critical exponents as a function of $N_{1}, N_{2}, N_{3}$ at the DFP, within the LPA 8 and LPA 8 with anomalous dimensions (LPA 8$+ \eta$). We do not exploit the scaling relations, but instead evaluate all critical exponents explicitly from the $\beta$ functions at the DFP. We check the scaling relation explicitly and give the deviation to this order of the truncation. The inclusion of the anomalous dimensions leads to a slightly larger violation of the scaling relation.}
\end{table}

{ We observe that the violation of the scaling relation is slightly larger when anomalous dimensions are taken into account. This is not necessarily surprising, as the main effect of a running wavefunction renormalization is \emph{not} to give a sizable improvement in the value of the critical exponents, but instead to provide a first reasonable estimate of the value of $\eta$ itself. We expect that an enlargement of our truncation, including a field-dependent wavefunction renormalization (or, in other words, momentum-dependent interaction terms), will improve the situation.} 

\begin{figure}[!t]
\includegraphics[width=0.8\linewidth]{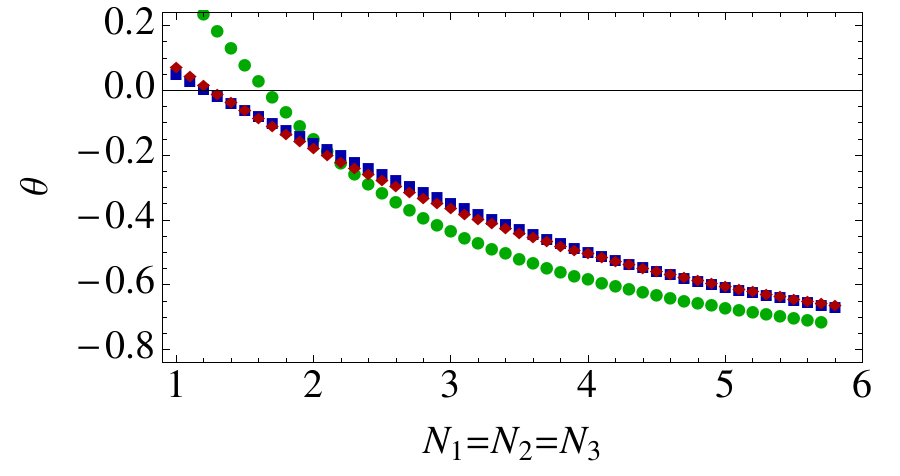}
\caption{\label{Fig:DFP}(Color online) We show the fourth-largest critical exponent at the DFP as a function of $N_{1} = N_{2} = N_{3}$, with the LPA 4 result (green circles), LPA 6 (blue squares) and LPA 8 (red diamonds). The LPA clearly converges rapidly.}
\end{figure}

Accordingly the stability of this FP can be determined completely from a knowledge of the Wilson-Fisher FP. Employing a LPA to order 12, including anomalous dimensions, we arrive at the results given in Tab.~\ref{Tab:DFPa}, cf.\ Ref.\ \cite{Eichhorn:2013zza}. Our results obtained within the LPA 8 for the three-field model show reasonable agreement with results deduced from the scaling relations, cf.\ Fig.\ \ref{Fig:DFP}. In fact, we may check the quantitative accuracy of the scaling relations explicitly, without referencing the results from the $O(N)$ model. We simply calculate the deviations $\Delta \theta_{4} = \theta_1 + \theta_2 - d - \theta_4$, $\Delta \theta_{5} = \theta_1 + \theta_3 - d - \theta_5$, and $\Delta \theta_{6} = \theta_2 + \theta_3 - d - \theta_6$, shown in Tab.\ \ref{Tab:DFPb} for the given data sets.

{Our explicit numerical results in LPA 8 deviate from the results inferred from the LPA 12 in the two-field model in some cases. There, we expect that an enlarged truncation in the three-field model will give results in full agreement with those deduced from the two-field case. As the number of couplings grows very substantially with the truncation, such an explicit check is beyond the scope of this work. Where LPA 8 and LPA 12 $+ \eta$ results deviate, the latter are more trustworthy.}

\subsection{Decoupled isotropic fixed point}
\label{SubSec:Decoupled isotropic fixed point}

For three interacting fields, we observe a new FP, where only one of the fields decouples, while the other two sectors show an enhancement of symmetry. For the following discussion, we will assume that it is the $\phi_1$-sector that decouples, while the remaining sectors have a $O(N_2+N_3)$ symmetry. Two other possible DIFPs exist, for which one of the other subsectors decouples, respectively.

Clearly, one positive critical exponent is inherited from the $O(N_1)$ symmetric and another one from the $O(N_2+N_3)$ symmetric Wilson-Fisher scaling spectrum. Two critical exponents that are relevant for the stability properties of this FP follow from the critical exponents of the spin-$1$ and spin-$2$ perturbations of the two-field isotropic $O(N_{2} + N_{3})$ FP, i.e., $y_{2,2}$ and $y_{4,4}$. While $y_{2,2}$ is always positive, $y_{4,4}$ becomes positive for $N_2 + N_3 > 2$. We therefore conclude, that the DIFP can only be stable for $N_2 = N_3 = 1$.

There are two further exponents that we need to consider to establish the stability of the DIFP solution. Both follow from a scaling relation exploiting the decoupling of the $1$-sector: At the FP, the operators $\mathcal{O}_1=\phi_1^2 (\phi_2^2+\phi_3^2)$ and $\mathcal{O}_2 = \phi_1^2 \left(\phi_2^2 - \frac{N_2}{N_2+N_3}(\phi_2^2+\phi_3^2) \right)$ correspond to eigendirections of the stability matrix. As we know from the Wilson-Fisher FP that the scaling dimension $[\phi_I^2] = -\frac{1}{\nu_I}+d$, we deduce that
\be
[\mathcal{O}_1] = -\frac{1}{\nu_1} - \frac{1}{\nu_{2+3}} + 2d, 
\ee
and accordingly the corresponding critical exponent is given by
\be
\theta_{6} = \frac{1}{\nu_{1}} + \frac{1}{\nu_{2+3}} - d ~.
\ee
Similarly, we deduce for the second operator that the corresponding critical exponent is given by
\be
\theta_{4} = \frac{1}{\nu_{1}} + y_{2,2} - d ~,
\ee
where $y_{2,2}$ is the scaling dimension of the coupling belonging to $\phi_2^2 - \frac{N_2}{N_2+N_3}(\phi_2^2+\phi_3^2)$ in the two-field case, cf.\ Ref.\ \cite{Eichhorn:2013zza}. The first relation is the one that arises for a two-field DFP, and gives a negative critical exponent (for values of $N_{1} > 1$).

At fixed $N_{2} = N_{3} = 1$, it is the critical exponent $\theta_{4}$ that decides about the stability of the FP. Using results from the two-field case (LPA 12 including an anomalous dimension, cf.\ Ref.\ \cite{Eichhorn:2013zza}) to obtain $\theta_{1} = y_{2,2}$, $\theta_{2} = \frac{1}{\nu_{1}}$, and $\theta_{3} = \frac{1}{\nu_{2+3}}$, we arrive at the results shown in Tab.~\ref{Tab:DIFP}. The DIFP is the stable FP for $N_{1} \geq 6$ and $N_{2} = N_{3} = 1$.

\begin{table}[!t]
\renewcommand{\arraystretch}{1.4}
\renewcommand{\tabcolsep}{4pt}
\begin{tabular}{c c c | c c c c c c}
$N_1$ & $N_2$ & $N_3$& $\theta_{1}$ & $\theta_{2}$ & $\theta_{3}$ & $\theta_{4}$ & $\theta_{5}$ & $\theta_{6}$ \\
\hline\hline
1 & 1 & 1 & 1.765 & 1.571 & 1.459 & 0.336 & -0.042 & 0.030 \\ 
2 & 1 & 1 & 1.765 & 1.459 & 1.459 & 0.224 & -0.042 & -0.092 \\
3 & 1 & 1 & 1.765 & 1.367 & 1.459 & 0.132 & -0.042 & -0.174 \\
4 & 1 & 1 & 1.765 & 1.296 & 1.459 & 0.061 & -0.042 & -0.245 \\
5 & 1 & 1 & 1.765 & 1.242& 1.459 & 0.007 & -0.042 & -0.299 \\
6 & 1 & 1 & 1.765 & 1.203 & 1.459 & -0.032 & -0.042 & -0.338 \\ 
\hline\hline
\end{tabular}
\caption{\label{Tab:DIFP}Critical exponents at the DIFP, using the LPA 12 including anomalous dimensions, and employing the above scaling relations.}
\end{table}

Note that the derivation of the scaling relations is based on the assumption that the operators corresponding to these couplings are eigenoperators of the stability matrix. This property is, to the best of our knowledge, an assumption in $d = 3$ \cite{Calabrese:2002bm}, and is usually not true within a truncation of the RG flow. Nevertheless, the stability properties are not incompatible with explicit numerical results within the LPA 8, where the transition to stability occurs already at $N_1=5$. { Explicitly, the critical exponents in LPA 8 at $N_1=5$ and $N_2=N_3=1$ read $\theta_1=1.783, \theta_2=1.193, \theta_3=1.399, \theta_4=-0.024,\theta_5=-0.027$ and $\theta_6=-0.395$.}

\vskip 5pt


\subsection{Decoupled biconical fixed point}
\label{SubSec:Decoupled biconical fixed point}

\begin{table}[!h]
\renewcommand{\arraystretch}{1.4}
\renewcommand{\tabcolsep}{2.7pt}
\begin{tabular}{c c c | c c c c c c}
$N_1$ & $N_2$ &$N_3$& $\theta_{1}$ & $\theta_{2}$ & $\theta_{3}$ & $\theta_{4}$ & $\theta_{5}$ & $\theta_{6}$ \\
\hline\hline
1 & 1.2 & 1 & {\bf 1.753} & {\bf 1.381} & \textit{1.537} & 0.285 & -0.075 & {\bf -0.005} \\
1 & 1.4 & 1 & {\bf 1.535} & {\bf 1.448} & \textit{1.537} & 0.105 & -0.015 & {\bf -0.010} \\
1 & 1.5 & 1 & {\bf 1.537} & {\bf 1.462} & \textit{1.537} & 0.068 & 0.001 & {\bf -0.001} \\ 
1 & 1.2 & 2 & {\bf 1.753} & {\bf 1.381} & \textit{1.399} & 0.161 & -0.200 & {\bf -0.005} \\
1 & 1.4 & 2 & {\bf 1.535} & {\bf 1.448} & \textit{1.399} & -0.020 & -0.140 & {\bf -0.010} \\
1 & 1.5 & 2 & {\bf 1.537} & {\bf 1.462} & \textit{1.399} & -0.057 & -0.124 & {\bf -0.001} \\
1.2 & 1 & 5 & {\bf 1.753} & {\bf 1.381} & \textit{1.193} & -0.051 & -0.413 & {\bf -0.005} \\
1.5 & 1 & 5 & {\bf 1.537} & {\bf 1.462} & \textit{1.193} & -0.270 & -0.338 & {\bf -0.001} \\
\hline\hline
\end{tabular}
\caption{\label{Tab:BFP}Critical exponents at the DBFP in the LPA 8. Critical exponents emphasized in \textbf{bold} font correspond to those of the two-field BFP, while those in \textit{italic} arise in the sector with $O(N_3)$ group symmetry. Here, we list the DBFP for values of the field components $N_{I}$, where it is characterized by $\Delta > 0$ (see text).}
\end{table}

Beyond the isotropic and decoupled FP solutions, the two-field models feature another scaling solution, which is the biconical FP \cite{PhysRevLett.33.813, Calabrese:2002bm, Folk:2008mi}. It is stable only in a restricted parameter region of these models, where $\left. \Delta^{\textrm{2-field}} \right|_{\textrm{BFP}} = \lambda_{2,0} \lambda_{0,2} - \lambda_{1,1}^2 > 0$, as it transfers the stability from the IFP to the DFP. Certainly, this FP should similarly manifest itself in the three-field case. While one of the three sectors decouples, the two remaining sectors should feature nondegenerate couplings, and we expect that in a given range of the parameter space such a decoupled BFP will be stable. 

{ To obtain as precise results as possible, we should make use of all methods available to us. In fact, results obtained using an $\epsilon$-expansion around $d=4$ in the two-field case, allow us to infer the stability of the decoupled biconical fixed point in the three field case in one important instance: }
From Refs.\ \cite{Folk:2008mi, Calabrese:2002bm}, we know that the biconical FP is stable for $N_1 = 1, N_2 = 2$ in two-field models (and similarly when the sectors are interchanged). Combined with the pattern in Tab.\ \ref{Tab:BFP} for the additional critical exponents in the three-field model, we conjecture that the DBFP is stable for $N_1 = 1, N_2 = 2, N_3\geq 2$ (up to a permutation of the three sectors). To calculate the corresponding critical exponents directly in the three-field model, we expect that an extended truncation will be necessary, taking into account a field-dependent wavefunction renormalization.


\subsection{Search for further stable fixed points}
\label{SubSec:Search for further stable fixed points}

\begin{figure}[!t]
\includegraphics[width=0.8\linewidth]{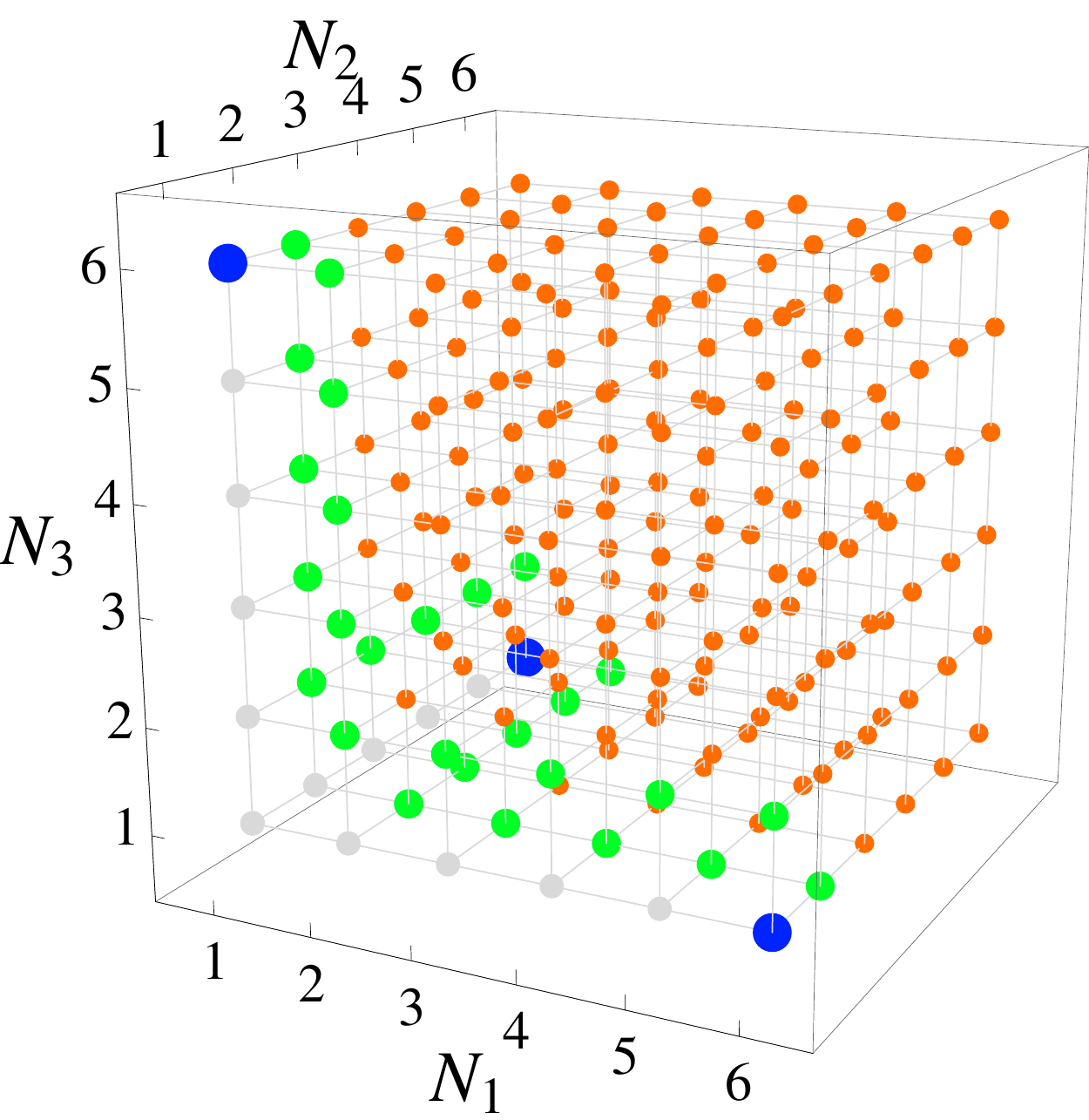}
\caption{\label{Fig:StabilityPlot}(Color online) We show the stable DIFP (large blue dots), the stable DFP (small orange dots) and points without a stable FP (small gray dots) using the LPA 12$+\eta$ results. We also include points where the DBFP is conjectured to be stable (middle-sized green dots).}
\end{figure}

We summarize our results obtained so far in Fig.~\ref{Fig:StabilityPlot}. The figure shows the stable FP solution for the corresponding values of field components, $(N_1, N_2, N_3)$. Apparently, no stable FP exists in the range $N_{1} < 6$, $N_{2} = N_{3}=1$ (up to a permutation of the sectors) that can be derived from the known scaling solutions in the one- and two-field models. This motivates an independent analysis of fully coupled FPs in the three-field model, which we describe in the following sections.

\subsubsection{Stability trading between stable FPs}
\label{SubSec:Stability trading between stable FPs}

\begin{figure}[!h]
\includegraphics[width=0.8\linewidth]{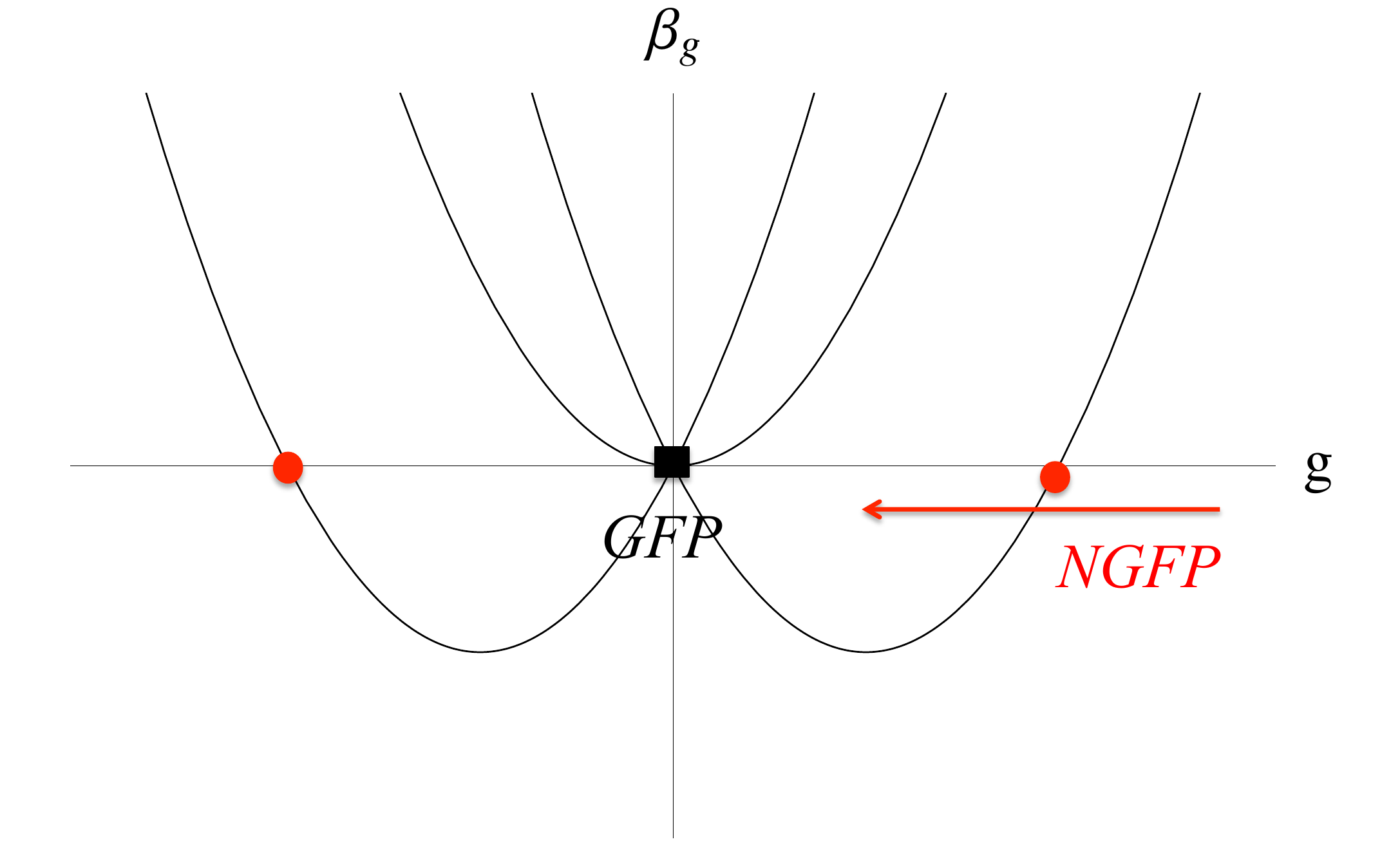}
\caption{\label{Fig:StabilityTrading}To illustrate the mechanism how stability properties are exchanged between FPs, we examine the $\beta$-function $\beta = g(c + g)$ for varying parameter $c$ describing the position of the nontrivial FP. As the FPs pass by one another, the local derivatives $\theta = -\beta'(g)$ exchange their sign indicating an interchange of their IR stability properties. }
\end{figure}

Generally, the $\beta$ functions are non-polynomial functions of a large number of couplings in the multifield models. In the local potential type approximations the number of parameters and couplings increases from 9 to 34 if the order of the truncation is changed from 4 to 8. Thus, finding FPs in the multifield models becomes a highly nontrivial search for zeros of the $\beta$ functions in a high-dimensional parameter space. In the following, we consider strategies to identify new FP solutions using a simple example. Consider the following $\beta$ function which is expanded in terms of the coupling $g$ (assuming that higher than quadratic terms are zero):
\be
\beta_g = g ( c + g ),
\ee
where $c$ is a function of the parameters of the model (e.g., dimensionality, number of field components, etc.) and possibly other couplings in a given truncation of the theory. Note that such a form captures the essential properties of typical fixed points, as it allows both for a trivial Gaussian FP and a nontrivial interacting FP, as a function of the parameter $c$. The critical exponent at a fixed point is given by
\be
\theta = - \left. \frac{\partial\beta_g}{\partial g} \right|_{\textrm{FP}} = \left. - c - 2 g \right|_{\rm FP}. 
\ee
Assuming that it is the exponent $\theta$ that decides about the stability of the FP, we may distinguish the following scenarios: For $c < 0$, the interacting FP at $g = -c$ is infrared stable, whereas the Gaussian FP is unstable. As the parameter $c$ increases towards positive values (as a function of, e.g., $N_I$) the two FPs will approach each other. At $c = 0$, both FPs collide and exchange their stability properties. Moving apart again for $c > 0$, the interacting FP has become the unstable one, whereas the noninteracting FP is stable, cf.\ Fig.\ \ref{Fig:StabilityTrading}. This simple example demonstrates that FPs will typically change their stability when they collide, as was also observed in Ref.\ \cite{Eichhorn:2013zza} in the two-field model with $O(N_{1})\oplus O(N_{2})$ symmetry.

Inspired by the above stability-trading mechanism
we may devise a strategy to identify new FP solutions. Our search for FPs will concentrate on the vicinity of points in coupling space, where a known FP loses its stability.


\subsubsection{Fully coupled fixed points in the three-field model}
\label{SubSec:Fully coupled fixed points in the three-field model}

It turns out that the three-field model works in a different way from the mechanism described above, which applies in the two-field case: Within the LPA 4, we observe that the IFP, DFP, and DIFP have partially overlapping stability regions. No similar behavior occurs in two-field models, where stability regions of different FPs always touch, but never overlap, due to the above stability-trading mechanism. However, this changes dramatically as we include anomalous dimensions: While the IFP inhabits the same points, the DIFP is now only stable for $N_1 \geq 4$, $N_2 = N_3=1$ etc.\ Extensive numerical searches did not reveal a stable FP for $N_1 =3$, $N_2 = N_3=1$, and similarly for the cases where the sectors are interchanged. A similar result holds for higher orders of the LPA: As there are a larger number of independent operators that serve as a basis for the LPA, and thus potentially relevant directions in the three-field model, the stability exchange mechanism may not be captured by the simple model considered above. To elucidate the differences in the three-field model, we will focus on results obtained within the LPA to 8th order. 

\begin{figure}[!t]
\includegraphics[width=\linewidth]{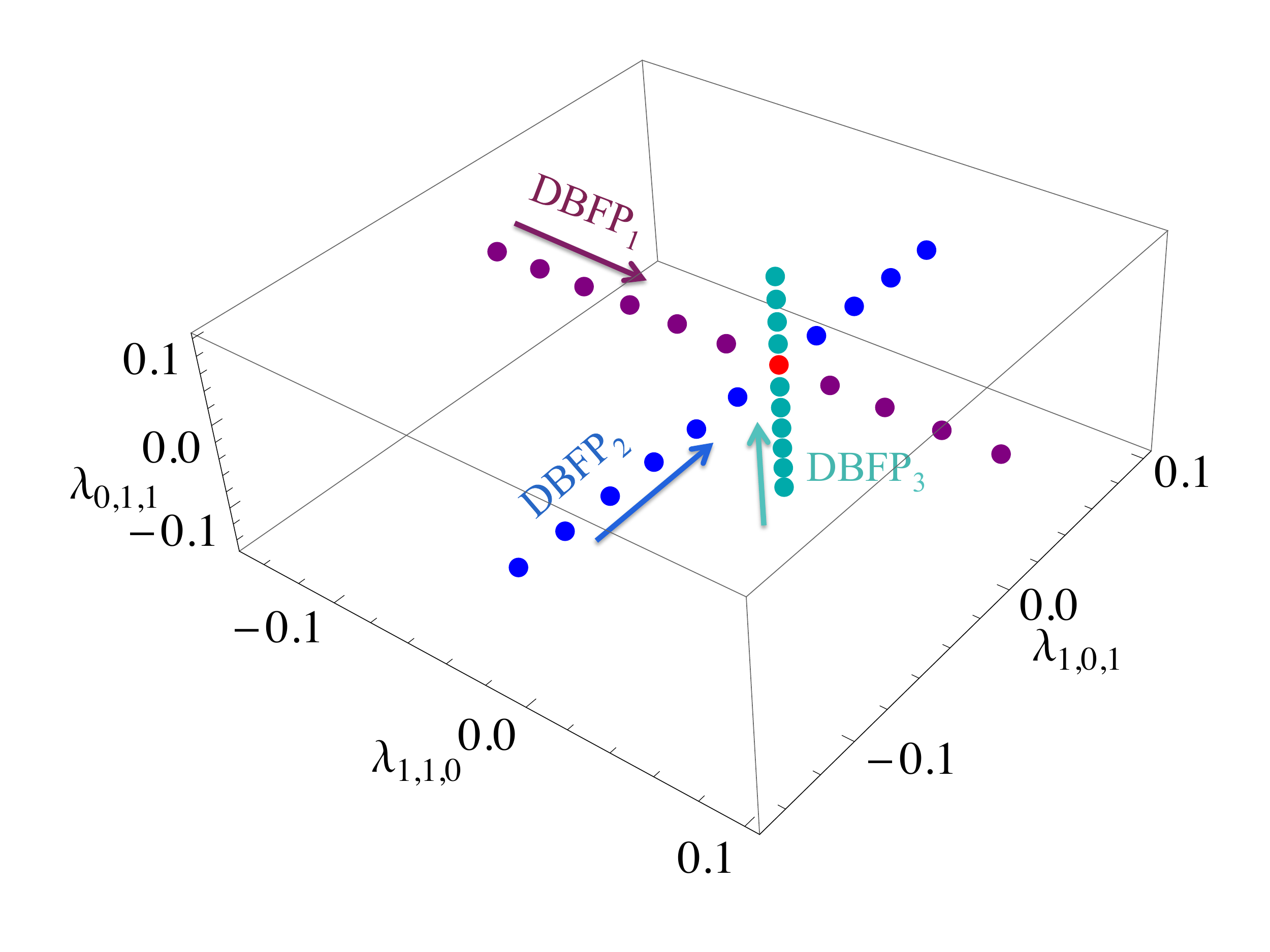}
\caption{\label{BFPDFPcollision}We plot $\lambda_{1,0,1}, \lambda_{0,1,1}$, and $\lambda_{1,1,0}$ as a function of $N_I$ for the three DBFPs. At $N_I=1.244$ they each pass through the origin of that coordinate system, where the DFP sits. For $N_I>1.244$, the three BFPs move away from each other towards more negative values of the mixed couplings.}
\end{figure}

\begin{figure}[!h]
\includegraphics[width=0.8\linewidth]{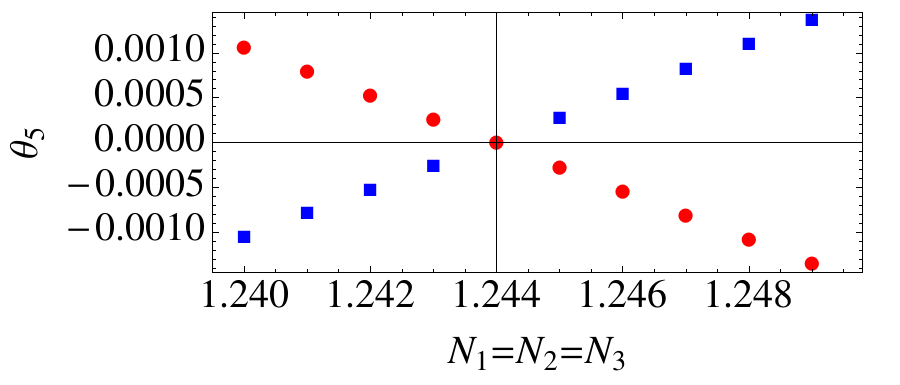}
\includegraphics[width=0.8\linewidth]{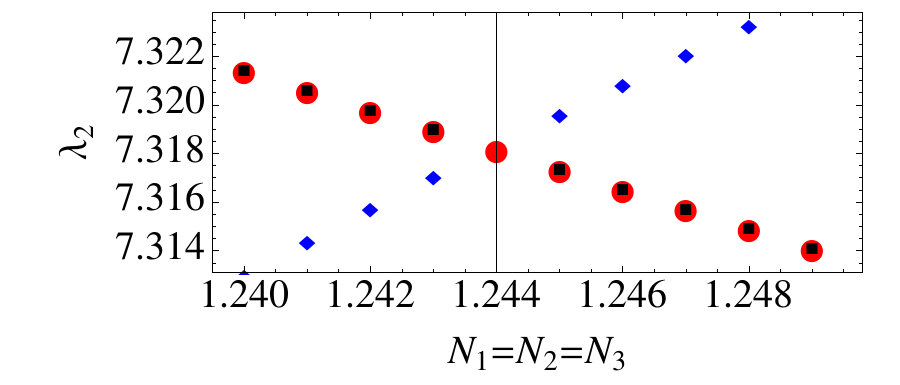}
\caption{\label{BFPDFPthetas}We plot the fifth-largest critical exponent at the DBFPs (blue points of increasing value) and the corresponding critical exponent at the DFP (red points of decreasing value) as a function of $N_I$ (upper panel). Below, we show the coordinates of the couplings $\lambda_{2,0,0}$ and $\lambda_{0,2,0}$ at the DBFP (blue diamonds and black squares) and the couplings $\lambda_{2,0,0}= \lambda_{0,2,0} = \lambda_{0,0,2}$ at the DFP (red dots) in the vicinity of the collision point.}
\end{figure}

As a first example, let us consider the point $N_1 = N_2 = N_3 \approx 1.25$. Here, the DFP is stable, but it gains three additional relevant directions around $N_1 = N_2 = N_3 \approx 1.244$. As within the two-field model, this point is marked by a collision with a decoupled biconical FP (DBFP). The main difference is that within a three-field model, three generalizations of the BFP exist, cf.\ Sec.\ \ref{SubSec:Decoupled biconical fixed point}. For $N_I =1.25$, $I = 1,2,3$, all of these FPs feature five relevant directions. Toward smaller $N_I$, all three DBFPs approach the DFP, and simultaneously collide with it at $N_I \approx 1.244$. At this point, the DFP gains one relevant direction from each of the three DBFPs, which subsequently feature only four relevant exponents, cf.\ Fig.\ \ref{BFPDFPcollision} and Fig.~\ref{BFPDFPthetas}. Thus this FP loses stability in a fixed-point collision. The central difference to the two-field case lies in the fact that the symmetry of the model which forces the DFP to collide with three other FPs simultaneously (for $N_1 = N_2 = N_3$ there is an exchange symmetry $N_1 \leftrightarrow N_2 \leftrightarrow N_3$). As each of them starts off with five relevant directions, the collision does not produce a stable FP for $N_I \leq 1.24$, but instead leaves behind three FPs with four relevant directions each.

\begin{figure}[!h]
\includegraphics[width=0.9\linewidth]{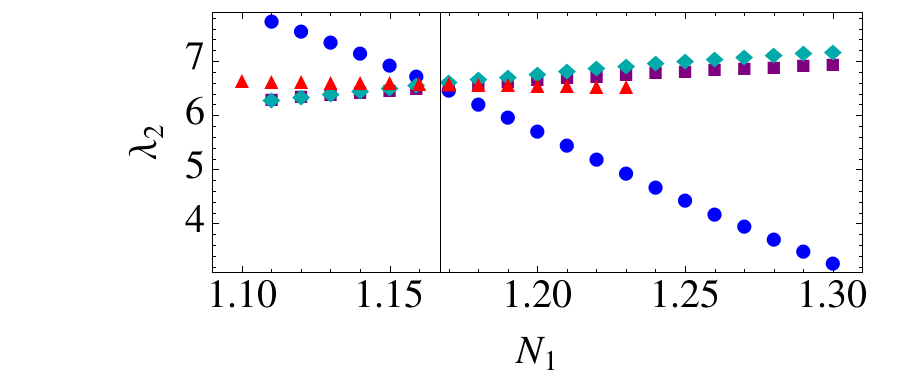}\\
\includegraphics[width=0.9\linewidth]{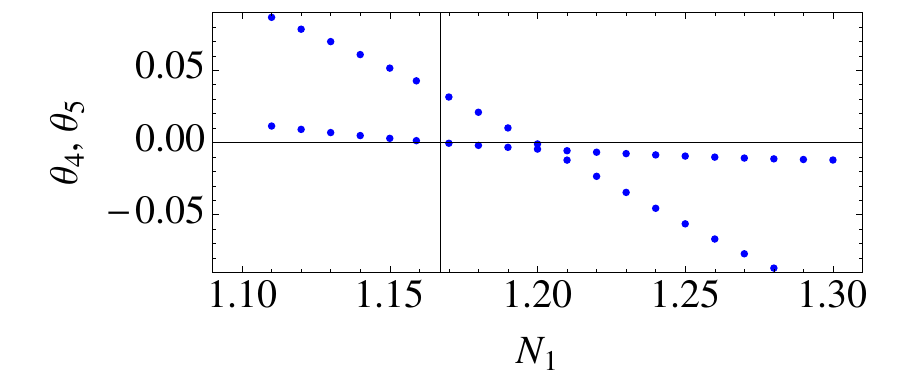}\\
\includegraphics[width=0.9\linewidth]{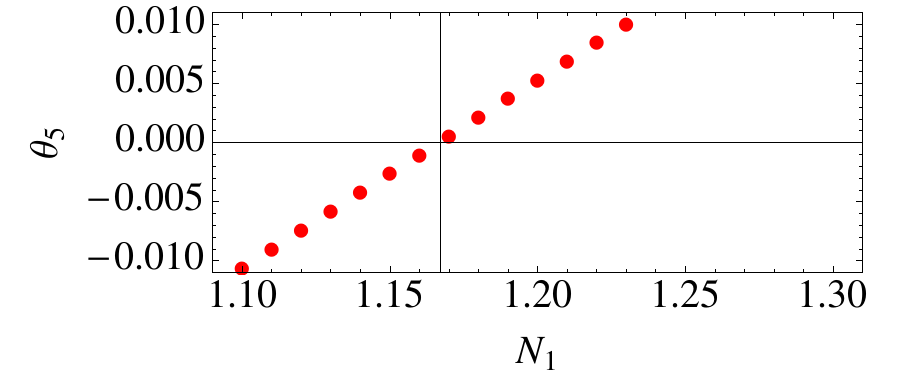}
\caption{\label{BFPDIFP}(Color online) Here, we show the couplings $\lambda_{2,0,0}$, $\lambda_{0,2,0}$, and $\lambda_{1,1,0}$ at the DBFP (blue circles, purple squares, cyan diamonds) and $\lambda_{2,0,0} = \lambda_{0,2,0} = \lambda_{1,1,0}$ (red triangles) at the DIFP (uppermost panel). We show the fourth- and fifth-largest critical exponent (middle panel) at the DBFP as a function of $N_1$ for $N_2=1, N_3=4$. Around $N_1 \approx 1.16$, the three couplings are clearly degenerate, as expected for a collision with the DIFP. At the same point, the fourth critical exponent crosses zero and becomes negative. Simultaneously, the fifth critical exponent at the DIFP (lower panel) crosses zero and becomes positive.}
\end{figure}

\begin{figure}[!h]
\includegraphics[width=\linewidth]{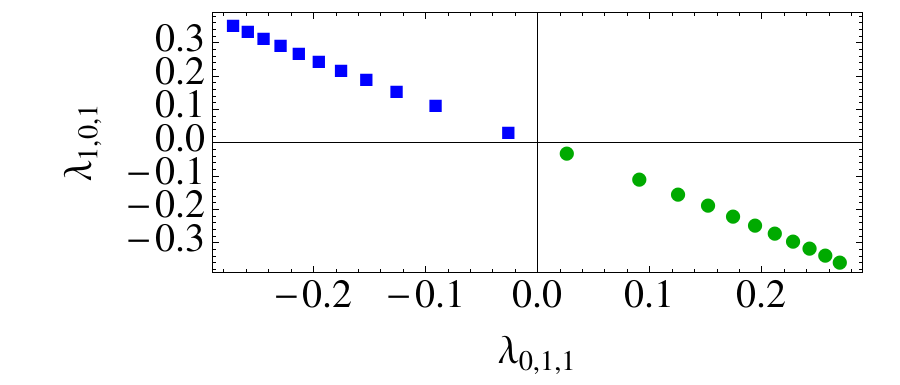}\\
\includegraphics[width=\linewidth]{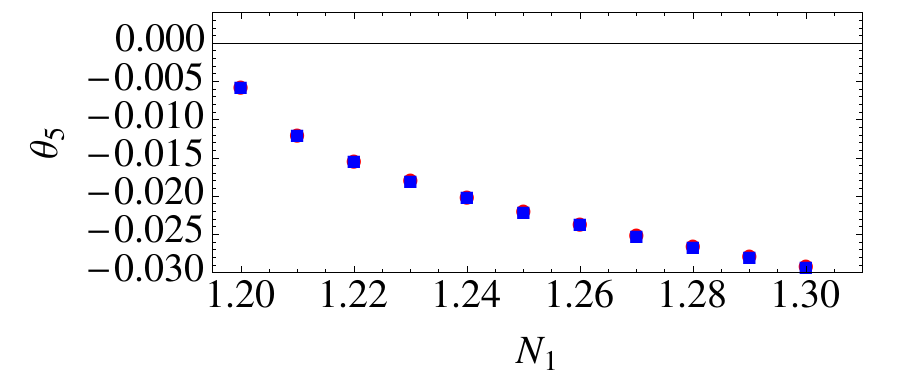}\\
\caption{\label{BFPcollision}Here, we show the couplings $\lambda_{1,0,1}$ and $\lambda_{0,1,1}$ at the two anisotropic FPs. As a function of $N_1$ they converge towards zero, which is where the DBFP is sitting. The collision occurs at $N_1 \approx 1.2$, which is where the fifth critical exponent of these two FPs (which numerically is nearly the same for both) approaches zero.}
\end{figure}

As a second example of new behavior in three-field models, we consider the point where the DBFP collides with the DIFP in the LPA 8. We fix $N_2 =1$ and $N_3=4$: Then the DBFP has five positive critical exponents at $N_1=1$. Going to larger values of $N_1$, it collides with the DIFP at $N_1 \approx 1.6$. During this collision, the DIFP becomes unstable, and the DBFP gains one negative critical exponent, cf.\ Fig.\ \ref{BFPDIFP}. Similar to the previous scenario, this FP collision is not sufficient to make the DBFP stable. Following the DBFP to even larger values of $N_1$, it undergoes another collision, this time being hit by two other FPs simultaneously. This is a novel feature that is not observed in simple two-field models. Starting in the region $N_1 < 1.2$, these FPs do not seem to exist for real values of the couplings -- at least no sign of them showed up in extensive numerical searches. They can be thought of as being created at the collision. Following this collision, they quickly move away from the collision point for increasing values of $N_{1}$. Each of the newly created FPs features four relevant critical exponents, while the DBFP is stable, cf.\ Fig.\ \ref{BFPcollision}. Both new FPs are anisotropic, and define a new universality class that occurs for the first time if three fields are coupled, cf.\ Tab.\ \ref{Tab:AFP}.

\begin{table}[!t]
\renewcommand{\arraystretch}{1.4}
\renewcommand{\tabcolsep}{2.7pt}
\begin{tabular}{ccc|ccc|ccc|c}
$N_1$ & $N_2$ & $N_3$ & $\lambda_{2,0,0}$ & $\lambda_{0,2,0}$ & $\lambda_{0,0,2}$ & $\lambda_{1,1,0}$ & $\lambda_{1,0,1}$ & $\lambda_{0,1,1}$ & $\lambda_{1,1,1}$\\ \hline \hline
2 & 2 & 2 & 6.2 & 6.4& 6.4 & 1.8 & 1.8 & -2.2 & -2.6\\
2 & 1 & 2 & 6.5 & 7.5 & 6.7 & 1.2 & 0.5 & -2.1 & -1.4\\
2 & 1 & 3 & 6.2 & 7.0 & 5.8 & 3.2 & 0.7 & -1.8 & -2.2\\
2 & 1 & 4 & 6.0 & 6.5 & 5.1 & 4.6 & 0.6 & -1.3 & -1.6\\ \hline \hline
\end{tabular}
\vskip 4pt
\renewcommand{\arraystretch}{1.4}
\renewcommand{\tabcolsep}{6.8pt}
\begin{tabular}{ccc|ccc|ccc}
{} & {} & {} & $\theta_1$ & $\theta_2$ & $\theta_3$ & $\theta_4$ & $\theta_5$ & $\theta_6$ \\ \hline \hline
2 & 2 & 2 & 1.70 & 1.33 & 1.29 & 0.20 & -0.17 & -0.24\\
2 & 1 & 2 & 1.62 & 1.40 & 1.35 & 0.08 & -0.05 & -0.23\\
2 & 1 & 3 & 1.71 & 1.34 & 1.28 & 0.15 & -0.10 & -0.30\\
2 & 1 & 4 & 1.77 & 1.32 & 1.23 & 0.16 & -0.07 & -0.35\\ \hline \hline
\end{tabular}
\caption{\label{Tab:AFP}We list selected FP values and critical exponents of the first anisotropic FP in the LPA 8, restricting ourselves to couplings at fourth order in the fields, and additionally giving the value of the coupling associated with the three-field operator $\sim \phi_1^2 \phi_2^2 \phi_3^2$.}
\end{table}

Due to the complicated dynamics of FPs in the parameter space of three-field models, it seems that these models do not necessarily feature a stable FP solution for all values of the model parameters. One might now wonder, why FPs of three-field models show such a disproportionate increase in the number of relevant directions for small values of $N_I$. While the Wilson-Fisher FP has one relevant direction, and two-field models always feature FPs with only two relevant directions, three-field models show FPs with more than three relevant directions. We conjecture that this is related to the fact that the transition from a two-field to a three-field model implies the existence of more than one additional new class of operators: While the transition from one to two fields only adds the mixed interactions $\sim \left( \phi_1^{2} \right)^{m} \left( \phi_2^{2} \right)^{n}$, the transition to three-field models features three different classes of mixed interactions $\sim \left( \phi_I^{2} \right)^{m} \left( \phi_J^{2} \right)^{n}$, $I,J = 1,2,3$, in addition to the new three-field couplings $\sim \left( \phi_1^{2}\right)^{l} \left( \phi_2^{2} \right)^{m} \left( \phi_3^{2} \right)^{n}$. These seem to play a particularly important role for small $N_I$ and imply that the IR scaling properties cannot be accounted for by only three free relevant parameters. Note that this does not imply that the corresponding additional couplings need to become relevant. Since these operators do not necessarily correspond to eigendirections of the RG, they may mix with other operators, and yield corrections to the scaling spectrum.

Applying our results to determine the properties of possible multicritical points in phase diagrams for systems with three competing order parameters, we may conclude that models with small $N_I$ will typically feature a first order, rather than a second order (multicritical) transition. In particular, this applies to the phenomenologically relevant model of three interacting $Z_{2}$-Ising fields, i.e., $Z_{2} \oplus Z_{2} \oplus Z_{2}$ symmetry.


\section{Summary and conclusions}
\label{Sec:Summary and conclusions}

Here, we present a renormalization group study of IR stable FP solutions in three-field models with $O(N_1) \oplus O(N_2) \oplus O(N_3)$ symmetry. Our main results regarding the existence of stable FPs are summarized in Fig.~\ref{Fig:StabilityPlot}. Models in this class exhibit FPs that generalize the Wilson-Fisher FP, falling into three distinct categories, each characterized by the degree of symmetry-enhancement. We find a decoupled FP, a partially isotropic FP solution, and a fully isotropic FP solution. Their scaling spectrum can be deduced partially by considering perturbations around the single- and two-field models with $O(N)$ and $O(N_{1})\oplus O(N_{2})$ symmetry, respectively. We proceed by deriving scaling relations between different critical exponents to discover the stability properties of nontrivial FPs in the three-field case. Apart from the generalized Wilson-Fisher scaling solutions, we identify a decoupled biconical FP whose scaling properties are partly inherited from the BFP in the $O(N_{1})\oplus O(N_{2})$ symmetric model. As a main result of this work, we find that these FPs all show a significantly larger number of relevant critical exponents than in the two-field case, in the region of small $N_1$, $N_2$, and $N_3$. We tentatively connect this result to the existence of a large number of mixed interactions, and further conjecture, that similar results will hold for models with $n>3$ interacting fields.

Summarizing our results, we find no IR stable FP for a small number of fields ($N_{1} < 6$, $N_{2} = N_{3} = 1$, up to permutations of the fields) in $d = 3$ dimensions. This result is certainly unexpected, as there is no evidence for similar behavior in coupled two-field models. While, in principle, we cannot exclude the possibility that stable FPs exist in that region of parameter space, we find no evidence for their existence in extensive numerical searches for FPs of the nonperturbative $\beta$ functions. The identification of FPs in the three-field field models with $O(N_{1})\oplus O(N_{2})\oplus O(N_{3})$ is in general a difficult problem, since the search has to proceed through a high-dimensional coupling space. Nevertheless, the understanding of basic stability transitions between different FPs serves as a guiding principle to single out possible candidates for nontrivial FPs. Quite generally, in coupled-field models, stability seems to be inherited from single mergers or collisions of different FPs. Searches around such stability transition points have not yielded any FP that carries over the stability properties from the IFP (stable at small, noninteger values of $N_{I}<1$) to the decoupled FP. This indicates that the dynamics of FPs in $O(N_{1})\oplus O(N_{2})\oplus O(N_{3})$ symmetric three-field, or general $\bigoplus_{I} O(N_{I})$ symmetric multifield models is very different from that encountered in the simpler $O(N_1)\oplus O(N_2)$-type models.

From these results, we may conclude that models with phenomenological relevance such as, e.g., the $Z_{2} \oplus Z_{2} \oplus Z_{2}$ symmetric model will not feature a multicritical point in its phase diagram. Certainly, it is challenging to find three parameters that are accessible experimentally, and may be tuned to the multicritical point. This would be necessary to quantify the scaling behavior close to the corresponding FP, or to show the absence of such a transition. Nevertheless, it is conceivable that in the context of ultracold atomic systems such a control of the system might be achievable \cite{doi:10.1146/annurev-conmatphys-070909-104059}. 

Finally, let us comment on the general applicability of our results to other systems of interest. The renormalization group flow equations are derived for general $d$ Euclidean dimensions and can be applied to $d = 2$, relevant for critical behavior of low-dimensional condensed-matter systems, and $d = 4$, for multifield models of inflation, as well as possible extensions of the Higgs sector of the standard model. We leave this for future work.

\begin{acknowledgements}
We thank P.~Calabrese for discussions.
Research at Perimeter Institute is supported by the Government of Canada through Industry Canada and by the Province of Ontario through the Ministry of Research and Innovation. This work is supported by DOE grant No.\ DE-FG0201ER41195. M.M.S. is supported by the grant ERC-AdG-290623 and DFG grant FOR 723. A.E. would like to thank the University of Heidelberg and the University of Illinois at Chicago for hospitality during the early stages of this work.
\end{acknowledgements}

\begin{appendix}
\label{Appendix}

\section{Functional renormalization group}
\label{Sec:Functional renormalization group}

The nonperturbative functional renormalization group \cite{Wetterich:1992yh,Berges:2000ew,Polonyi:2001se,Pawlowski:2005xe,Gies:2006wv,Delamotte:2007pf} defines a functional flow for the scale dependent effective action $\Gamma_{k}$, which interpolates between the microscopic action $S$ defined at some ultraviolet cutoff scale $k = \Lambda$ and the full effective action $\Gamma_{k\rightarrow 0} = \Gamma$, when the renormalization group parameter $k$ is removed. It is given by
\be
\partial_t \Gamma_{k} = \frac{1}{2} \Tr \int\! \frac{d^{d}q}{(2\pi)^{d}} \left(\Gamma_k^{(2)} (q) + R_k(q) \right)^{-1} \partial_t R_k(q) ~,
\label{Eq:WetterichEquation}
\ee
where the logarithmic scale derivative is written in terms of the parameter $t = \ln (k/\Lambda)$, and the second functional derivative $\Gamma_k^{(2)}(p,q) = \frac{\delta^2 \Gamma_k}{\delta \chi(-p) \delta \chi(q)}$, and $\Gamma^{(2)}(q) (2\pi)^{d} \delta^{(d)}(p-q) \equiv \Gamma^{(2)} (p,q)$. Here, $\chi$ denotes the complete field content of our model and the trace $\Tr$ denotes a summation over internal degrees of freedom, i.e., both fields and field components. The regulator function $R_{k}$ implements a mass-like cutoff and regulates the infrared divergences. We take $R_{I J} (q) = R_{I}(q) \delta_{I J}$, $I, J = 1,2,3$, while the momentum dependence is given by $R_{I}(q) = Z_{I} (k^{2} - q^{2}) \theta(k^{2} - q^{2})$, where the wavefunction renormalization is scale dependent. This choice is referred to as the optimized regulator \cite{Litim:2000ci, Litim:2001up} which allows us to derive fully analytic expressions for the nonperturbative $\beta$ functions, and it is thus a convenient choice to identify possible scaling solutions.

\onecolumngrid
\section{Scale dependent effective potential}
\label{Sec:Scale dependent effective potential}

We derive the renormalization group flow equation for the effective potential by plugging our \textit{ansatz} Eq.\ \eqref{Eq:EffectivePotential} into the flow equation \eqref{Eq:WetterichEquation} and projecting the right-hand-side onto a constant field configuration:
\bea
\partial_t u_k &=& -d u_k +\sum_I (d-2+ \eta_I) \rho_I \partial_{\rho_{I}} u_{k} + { 2 v_{d} \sum_{I} \Big\{ (N_I-1) l_{0}^{(I)} \left( \partial_{\rho_{I}} u_{k} \right)} \nonumber\\ && { +\: l_{R}^{(I)} \left( \left\{ \partial_{\rho_{J}} u_{k} + 2 \rho_{J} \partial_{\rho_{J}}^{2} u_{k} \right\}, \left\{ 4 \rho_{J} \rho_{K} \partial_{\rho_{J}} \partial_{\rho_{K}} u_{k} \right\} \right)} \Big\} ~. 
\label{Eq:FlowEquationPotential}
\eea
Here, $v_{d}^{-1} = 2^{d+1} \pi^{\frac{d}{2}} \Gamma\left(\frac{d}{2}\right)$ arises from the volume integration, and the anomalous dimensions are defined as $\eta_I = - \partial_t \ln Z_I$. In the following the notation follows Sec.\ \ref{Sec:Model}. The threshold functions $l_{0}^{(I)}$ and $l_{R}^{(I)}$ (see, e.g., \cite{Braun:2011pp}) define the diagrammatic contributions to the renormalization group flow of the scale dependent effective potential, where the upper index indicates the corresponding sector. Using an optimized regulator function \cite{Litim:2000ci,*Litim:2001up} the threshold functions take the following form:
\begin{eqnarray}
l_{0}^{(I)}(w_{I}) &=& \frac{2}{d} \left( 1 - \frac{\eta_{I}}{d+2}\right)\frac{1}{1+w_{I}} ~, \\
\hspace{-20pt} l_{R}^{(1)}(\{ w_{I} \}, \{ \delta_{I,J}^{2} \}) &=& \frac{2}{d} \left( 1 - \frac{\eta_{1}}{d+2}\right) \frac{\left( 1 + w_{2} \right) \left( 1 + w_{3} \right) - \delta_{2,3}^{2}}{2 \delta_{1,2} \delta_{1,3} \delta_{2,3} 
- \delta_{1,2}^{2} \left( 1 + w_{3} \right) - \delta_{1,3}^{2} \left( 1 + w_{2} \right) - \delta_{2,3}^{2} \left( 1 + w_{1} \right)
+ \prod_{J} \left( 1 + w_{J} \right)} ~, \\
\hspace{-20pt} l_{R}^{(2)}(\{ w_{I} \}, \{ \delta_{I,J}^{2} \}) &=& \frac{2}{d} \left( 1 - \frac{\eta_{2}}{d+2}\right) 
\frac{\left( 1 + w_{1} \right) \left( 1 + w_{3} \right) - \delta_{1,3}^{2}}{2 \delta_{1,2} \delta_{1,3} \delta_{2,3} 
- \delta_{1,2}^{2} \left( 1 + w_{3} \right) - \delta_{1,3}^{2} \left( 1 + w_{2} \right) - \delta_{2,3}^{2} \left( 1 + w_{1} \right)
+ \prod_{J} \left( 1 + w_{J} \right)} ~, \\
\hspace{-20pt} l_{R}^{(3)}(\{ w_{I} \}, \{ \delta_{I,J}^{2} \}) &=& \frac{2}{d} \left( 1 - \frac{\eta_{3}}{d+2}\right) 
\frac{\left( 1 + w_{1} \right) \left( 1 + w_{2} \right) - \delta_{1,2}^{2}}{2 \delta_{1,2} \delta_{1,3} \delta_{2,3} 
- \delta_{1,2}^{2} \left( 1 + w_{3} \right) - \delta_{1,3}^{2} \left( 1 + w_{2} \right) - \delta_{2,3}^{2} \left( 1 + w_{1} \right)
+ \prod_{J} \left( 1 + w_{J} \right)} ~.
\end{eqnarray}
Note, that the nature of the interactions is such, that only the radial modes $l_{R}^{(I)}$ are affected by the couplings, e.g., $\delta_{1,2}^{2} = 4 \kappa_{1} \kappa_{2} \lambda_{1,1,0}^{2}$, and equivalent couplings between the remaining sectors. One may easily check that in the limit of vanishing couplings, $\delta_{1,3} \rightarrow 0$ and $\delta_{2,3} \rightarrow 0$, the radial contributions $l_{R}^{(I)}$ in the $(1,2)$ sectors reduce to the threshold functions that were already derived in the two-field model \cite{Eichhorn:2013zza}:
\begin{eqnarray}
\hspace{-20pt} l_{0}^{(I)}( w_{I} ) &=& \frac{2}{d} \left( 1 - \frac{\eta_{I}}{d+2}\right)\frac{1}{1+w_{I}} ~, 
\label{Eq:ThresholdFunctionL0} \\
\hspace{-20pt} l_{R}^{(1)}(\{ w_{I} \}, \{ \delta_{I,J}^{2} \}) &=& \frac{2}{d} \left( 1 - \frac{\eta_{1}}{d+2}\right) \frac{1 + w_{2}}{\left( 1 + w_{1} \right) \left( 1 + w_{2} \right)- \delta_{1,2}^{2}} ~, \\
\hspace{-20pt} l_{R}^{(2)}(\{ w_{I} \}, \{ \delta_{I,J}^{2} \}) &=& \frac{2}{d} \left( 1 - \frac{\eta_{2}}{d+2}\right) \frac{1 + w_{1}}{\left( 1 + w_{1} \right) \left( 1 + w_{2} \right)- \delta_{1,2}^{2}} ~.
\label{Eq:ThresholdFunctionLTildeR}
\end{eqnarray}
From \eqref{Eq:FlowEquationPotential} the flow equations for the couplings are derived by the differentiation with respect to the fields and successive projection onto a nonvanishing constant background field configuration $\rho_{I} = \kappa_{I}$, defined by the minimum of the effective potential. For some FPs it might be necessary to employ an expansion point where one or several of the $\kappa_{I}$ are vanishing. In this case $\beta_{\kappa_{3}} \equiv 0$. For a detailed discussion of this issue, we refer to Ref.\ \cite{Eichhorn:2013zza}. 

We obtain the $\beta$ functions for the couplings $\lambda_{l,m,n}$, $l + m + n \geq 2$:
\be
\beta_{\lambda_{l,m,n}} = \left. \left(\frac{\partial^{l+m+n}}{\partial \rho_1^{l} \partial \rho_2^{m} \partial \rho_3^{n}} \partial_t u_k + \beta_{\kappa_1} \frac{\partial^{(l+1)+m+n} u_k}{\partial \rho_{1}^{l+1} \partial \rho_2^m \partial \rho_3^n} + \beta_{\kappa_2} \frac{\partial^{l+(m+1)+n} u_k}{\partial \rho_{1}^{l} \partial \rho_2^{m+1} \partial \rho_3^n} + \beta_{\kappa_3} \frac{\partial^{l+m+(n+1)} u_k}{\partial \rho_{1}^{l} \partial \rho_2^m \partial \rho_3^{n+1}} \right) \right|_{\rho_I = \kappa_I} ~,
\ee
where $\beta_{\lambda_{l,m,n}} \equiv \partial_{t} \lambda_{l,m,n}$. The $\beta$ functions for the scale dependent dimensionless field expectation values $\kappa_{I}$, $I = 1,2,3$, are given by
\bea
\beta_{\kappa_1} &=& \frac{1}{\Delta} \left\{
- \Delta_{2,3} \partial_{\rho_1} \partial_t u_k + \left(\lambda_{1,1,0} \lambda_{0,0,2} - \lambda_{1,0,1} \lambda_{0,1,1}\right) \partial_{\rho_2} \partial_t u_k \right. \nonumber \\ && \left.\left. +\: \left(\lambda_{1,0,1} \lambda_{0,2,0} - \lambda_{1,1,0} \lambda_{0,1,1}\right) 
\partial_{\rho_3} \partial_t u_k \right\} \right|_{\rho_I = \kappa_I} ~, 
\eea
\bea
\beta_{\kappa_2} &=& \frac{1}{\Delta} \left\{ - \Delta_{1,3} \partial_{\rho_2} \partial_t u_k + \left( \lambda_{2,0,0} \lambda_{0,1,1} - \lambda_{1,1,0} \lambda_{1,0,1} \right) \partial_{\rho_3} \partial_t u_k \right. \nonumber\\ 
&& \left.\left. +\: \left( \lambda_{0,0,2} \lambda_{1,1,0} - \lambda_{1,0,1} \lambda_{0,1,1} \right) \partial_{\rho_1} \partial_t u_k \right\} \right|_{\rho_I = \kappa_I} ~, \\
\beta_{\kappa_3} &=& \frac{1}{\Delta} \left\{
- \Delta_{1,2} \partial_{\rho_3} \partial_t u_k  + \left( \lambda_{2,0,0} \lambda_{0,1,1} - \lambda_{1,1,0} \lambda_{1,0,1} \right) \partial_{\rho_2} \partial_t u_k \right. \nonumber\\ && \left.\left. +\: \left( \lambda_{0,2,0} \lambda_{1,0,1} - \lambda_{1,1,0} \lambda_{0,1,1}\right) \partial_{\rho_1} \partial_t u_k \right\} \right|_{\rho_I = \kappa_I} ~. 
\eea
Here, we have defined the coupling parameter
\be
\Delta_{1,2} = \lambda_{2,0,0} \lambda_{0,2,0} - \lambda_{1,1,0}^{2} ~,
\ee
and equivalently $\Delta_{1,3}$ and $\Delta_{2,3}$ (defined from the remaining quartic couplings in the three-field model), as well as the parameter
\bea
\Delta &=& - 2 \left( \lambda_{2,0,0} \lambda_{0,2,0} \lambda_{0,0,2} - \lambda_{1,1,0} \lambda_{1,0,1} \lambda_{0,1,1} \right) + \lambda_{0,0,2} \Delta_{1,2} + \lambda_{0,2,0} \Delta_{1,3} + \lambda_{2,0,0} \Delta_{2,3} ~.
\eea
These parameters quantify the symmetry enhancement properties of the system. In particular, for certain symmetry enhanced FPs, these quantities vanish exactly.

\section{Wavefunction renormalization and anomalous dimensions}
\label{Sec:Wavefunction renormalization and anomalous dimensions}

To determine the scale dependence of the field independent renormalization factor $Z_{I}$ from the functional flow equation \eqref{Eq:WetterichEquation} we perform a projection of the flow onto operators of the type $\sim \left( \partial_{\mu} \phi_{I} \right)^{2} $. This yields the scale dependence of the coefficient for the corresponding operator in the effective action. We have
\begin{equation}
\partial_{t} Z_{I} = \frac{(2\pi)^{d}}{\delta^{(d)}(0)} \lim_{Q\rightarrow 0} \frac{\partial}{\partial Q^{2}} \frac{\delta^{2}}{\delta \varphi_{I}(-Q) \delta \varphi_{I}(Q)} \partial_{t} \Gamma_{k} ~,
\label{Eq:RenormalizationFactorProjection}
\end{equation}
where the functional derivatives are taken with respect to the Nambu-Goldstone (NG) degrees of freedom $\varphi_{I}$ in the $I$-sector. Note, there is no summation implied over the $I$-index. For details of the derivation in the context of the $O(N)$ vector model, we refer to \cite{Tetradis:1993ts,Berges:2000ew}. 

The anomalous dimensions are defined via the scaling contribution to the wavefunction renormalization, i.e., $\eta_{I} = - \partial_{t} \ln Z_{I}$, and take the following form in the three-field model:
\begin{eqnarray}
\eta_{1} &=& \frac{16 v_{d}}{d} \Xi^{-1} \left\{ \kappa_{2} \lambda_{1,1,0}^2 + \kappa_{3} \lambda_{1,0,1}^2 + \kappa_{1} \left( \lambda_{2,0,0} + 2 \kappa_{2} \Delta_{1,2} + 2 \kappa_{3} \Delta_{1,3} + 4 \kappa_{2} \kappa_{3} \Delta \right)^2 \right. \nonumber\\ && \left. +\: 4 \kappa_{2} \kappa_{3} \left( \Delta_{2,3} \left( \lambda_{2,0,0} + \kappa_{2} \Delta_{1,2} + \kappa_{3} \Delta_{1,3} \right) - \Delta \left(1 + \kappa_{2} \lambda_{0,2,0} + \kappa_{3} \lambda_{0,0,2} \right) \right) \right\} ~, 
\label{Eq:Eta1} \\
\eta_{2} &=& \frac{16 v_{d}}{d} \Xi^{-1} \left\{ \kappa_{1} \lambda_{1,1,0}^{2} + \kappa_{3} \lambda_{0,1,1}^2 + \kappa_{2} \left( \lambda_{0,2,0} + 2 \kappa_{1} \Delta_{1,2} + 2 \kappa_{3} \Delta_{2,3} + 4 \kappa_{1} \kappa_{3} \Delta \right)^2 \right. \nonumber\\ 
&& \left. +\: 4 \kappa_{1} \kappa_{3} \left( \Delta_{1,3} \left( \lambda_{0,2,0} + \kappa_{1} \Delta_{1,2} + \kappa_{3} \Delta_{2,3} \right) - \Delta \left(1 + \lambda_{2,0,0} \kappa_{1} + \lambda_{0,0,2} \kappa_{3} \right) \right) \right\} ~, 
\label{Eq:Eta2} \\
\eta_{3} &=& \frac{16 v_{d}}{d} \Xi^{-1} \left\{ \kappa_{1} \lambda_{1,0,1}^2 + \kappa_{2} \lambda_{0,1,1}^2 + \kappa_{3} \left( \lambda_{0,0,2} + 2 \kappa_{1} \Delta_{1,3} + 2 \kappa_{2} \Delta_{2,3} + 4 \kappa_{1} \kappa_{2} \Delta \right)^2 \right. \nonumber\\ 
&& \left. +\: 4 \kappa_{1} \kappa_{2} \left( \Delta_{1,2} \left( \lambda_{0,0,2} + \kappa_{1} \Delta_{1,3} + \kappa_{2} \Delta_{2,3} \right) - \Delta \left(1 + \lambda_{2,0,0} \kappa_{1} + \lambda_{0,2,0} \kappa_{2} \right) \right) \right\} ~,
\label{Eq:Eta3}
\end{eqnarray} 
where the prefactor in the above expressions is defined as:
\begin{eqnarray}
\Xi &=& \Big(1 + 2 \sum_{I} \lambda_{I} \kappa_{I} + 4 \sum_{I < J} \kappa_{I} \kappa_{J} \Delta_{I,J} + 8 \kappa_{1} \kappa_{2} \kappa_{3} \Delta \Big)^2 ~.
\end{eqnarray}
Note, that Eqs.\ \eqref{Eq:Eta1} -- \eqref{Eq:Eta3} reduce to the two-field results in the limit $\delta_{1,3}^{2} = \delta_{2,3}^{2} = 0$ (cf.\ Appendix \ref{Sec:Scale dependent effective potential})
\begin{eqnarray}
\hspace{-20pt} \eta_{1} &=& \frac{16 v_{d}}{d} \frac{ \kappa_{2} \lambda_{1,1}^2 + \kappa_{1} \left(\lambda_{2,0} + 2 \kappa_{2} \Delta_{1,2} \right)^2}{\left(1 + 2 \kappa_{1} \lambda_{2,0} + 2 \kappa_{2} \lambda_{0,2} + 4 \kappa_{1} \kappa_{2} \Delta_{1,2} \right)^2} ~, \\
\hspace{-20pt} \eta_{2} &=& \frac{16 v_{d}}{d} \frac{ \kappa_{1} \lambda_{1,1}^2 + \kappa_{2} \left(\lambda_{0,2} + 2 \kappa_{1} \Delta_{1,2} \right)^2}{\left(1 + 2 \kappa_{1} \lambda_{2,0} + 2 \kappa_{2} \lambda_{0,2} + 4 \kappa_{1} \kappa_{2} \Delta_{1,2} \right)^2} ~,
\end{eqnarray}
where we have written $\lambda_{1,1} \equiv \lambda_{1,1,0}$, etc., as the $3$-sector effectively decouples in this case. From here, it is easy to identify the decoupling and symmetry enhancement scenarios for the spectrum of anomalous dimensions, by considering the proper limits of the couplings between different sectors. These results for the two-field case are equivalent to the anomalous dimensions given in Refs.\ \cite{Friederich:2010hr,Wetzel:2013}.
\twocolumngrid

\end{appendix}

\bibliography{references}

\end{document}